\documentclass[aps,prd,onecolumn,eqsecnum,amsmath,nofootinbib,preprintnumbers]{revtex4}%

\usepackage{etoolbox} 
\usepackage{color,graphicx,float}
\usepackage{amsfonts,amssymb,theorem,mathrsfs,times}
\usepackage{bm}
\usepackage{mathtools}
\usepackage{amsfonts,amssymb,theorem,mathrsfs}
\usepackage{dsfont} 
\usepackage{setspace}
\usepackage{ulem}
\usepackage[colorlinks,linkcolor=blue,anchorcolor=blue,citecolor=blue]{hyperref}
\usepackage{cancel}  
\usepackage{booktabs}
\usepackage{multirow}

{\theorembodyfont{\upshape}
}
{\theorembodyfont{\upshape}
}
{\theorembodyfont{\upshape}
}
{\theorembodyfont{\upshape}
}
{\theorembodyfont{\upshape}
}
{\theorembodyfont{\upshape}
}

\newcommand{\dalm}{\kern1pt\vbox{\hrule height 0.9pt\hbox{\vrule width
0.9pt\hskip 2.5pt\vbox{\vskip 5.5pt}\hskip 3pt\vrule width
0.3pt}\hrule height 0.3pt}\kern1pt}

\begin{document}
\preprint{\hfill {\small{USTC-ICTS/PCFT-24-39}}}

\title{\textbf{Image of Quantum Improved Regular Kerr Black Hole and Parameter Constraints from EHT Observations  
}}

%

\author{ Li-Ming Cao$^{a\, ,b}$\footnote{e-mail
address: caolm@ustc.edu.cn}}

\author{ Long-Yue Li$^c$\footnote{e-mail
address: lily26@mail.ustc.edu.cn}}

\author{ Xia-Yuan Liu$^b$ \footnote{e-mail
address: liuxiayuan@mail.ustc.edu.cn }}

\affiliation{$^a$Peng Huanwu Center for Fundamental Theory, Hefei, Anhui 230026, China}

\affiliation{${}^b$
Interdisciplinary Center for Theoretical Study and Department of Modern Physics,\\
University of Science and Technology of China, Hefei, Anhui 230026,
China}

\affiliation{${}^c$
Institute of Fundamental Physics and Quantum Technology, Department of Physics, School of Physical Science and Technology,
Ningbo University, Ningbo, Zhejiang 315211, China}

\begin{abstract}
    Quantum Improved Regular Kerr (QIRK) black hole is a rotating regular black hole model constructed based on the asymptotic safety method. 
    The model eliminates the ring singularity and prevents the formation of closed timelike curves, 
    while retaining well-defined thermodynamic properties.  
    Given these properties, 
    probing the observable features of the QIRK black hole is important. 
    In this work, we numerically determine the region of parameter space in which the QIRK spacetime remains regular, admits an event horizon, and is free of closed timelike curves.  
    Subsequently, 
    we simulate images of a QIRK black hole surrounded by a thin accretion disk. We find the primary effect of the quantum correction parameter,
    \(\widetilde{\omega}\),
    is a systematic reduction in the overall observed intensity, with only subtle effects on the image geometry. 
    Using observational data from the Event Horizon Telescope (EHT) for Sgr A* and M87*, 
    we further constrain the parameters of the QIRK black hole. 
    Moreover, since there exist QIRK parameters that are free of singularities and can admit closed timelike curves, 
    we investigate the images of CTCs under these conditions.
    These results reveal the distinctive observational features of the QIRK spacetime
    and provide a quantitative basis for assessing its viability as an astrophysical candidate.
\end{abstract}

\maketitle
	
\author{ Xia-Yuan Liu$^b$\footnote{e-mail
		address: liuxiayuan@mail.ustc.edu.cn}}	


\date{\today}
	

\section{Introduction}

Black holes, predicted by general relativity, 
are one of the most enigmatic objects in the universe. 
Classical black holes, such as the Kerr black hole
\cite{kerr_gravitational_1963}
which describes a rotating black hole, play a crucial role in understanding numerous astrophysical phenomena, including quasars and gamma-ray bursts. 
However, 
a significant issue with these classical solutions is the presence of singularities, 
characterized by the divergence of curvature invariants and geodesic incompleteness
\cite{wald_general_1984,hawking_large_1973}. 
To address this problem, regular black holes (RBHs) were introduced. 
Unlike traditional black holes, 
they are free from essential singularities throughout the entire spacetime 
\cite{bardeen_non-singular_1968,
dymnikova_vacuum_1992,ayon-beato_regular_1998,bronnikov_regular_2001}. 
The first regular black hole model was proposed by Bardeen
\cite{bardeen_non-singular_1968}, 
and was later interpreted as a solution derived from Einstein's field equations with nonlinear electromagnetic field
\cite{Ayon-Beato:2000mjt}. 
This approach has been extended to describe various spherically symmetric RBH models 
\cite{Ayon-Beato:2000mjt,
PhysRevLett.96.031103,
bronnikov_regular_2001,
fan_construction_2016,
bronnikov_regular_2006,
bronnikov_field_2022}. 
Since Bardeen's work, numerous other regular black hole models have been developed 
\cite{Ayon-Beato:1999kuh,
burinskii_new_2002,
dymnikova_regular_2004,
berej_regular_2006,
junior_regular_2015,
sajadi_nonlinear_2017,
toshmatov_comment_2018,
bambi_rotating_2013,
toshmatov_rotating_2014,
azreg-ainou_generating_2014,
ghosh_nonsingular_2015,
ghosh_radiating_2015}, 
including several rotating RBH models
\cite{bambi_rotating_2013,
toshmatov_rotating_2014,
azreg-ainou_generating_2014,
ghosh_nonsingular_2015,
ghosh_radiating_2015}.

Two primary methods have been established for constructing regular black hole models. 
The first one involves solving Einstein's field equations with specific matter sources 
\cite{noauthor_noncommutative_2006,
noauthor_non-commutative_2009,
noauthor_non-minimal_2007}. 
The second method derives RBHs as quantum corrections to classical black holes with singularities. The notable examples come from frameworks such as loop quantum gravity\cite{modesto_disappearance_2004,
gambini_black_2008,perez_black_2017,
Battista:2023iyu}.
A promising alternative is the asymptotic safety scenario 
\cite{Reuter:1996cp,
Souma:1999at,
doi:10.1142/10369,2019qgfr.book.....R,bonanno_renormalization_2000,
koch_black_2014,
chen2024quantumimprovedregularkerr}, 
which suggests that quantum gravity effects, particularly through the functional renormalization group, 
introduce a repulsive force near black hole cores
\cite{bonanno_renormalization_2000,Bonanno:2006eu}, 
potentially resolving singularities. 
Many such approaches have been studied for various black hole solutions
\cite{reuter_running_2004, 
falls_black_2012, 
platania_black_2023, 
eichhorn_black_2022, 
koch_structural_2013, 
pawlowski_quantum-improved_2018, 
ishibashi_quantum_2021, 
Litim:2013gga, 
Harst:2011zx, 
Falls:2012nd, 
Adeifeoba:2018ydh, 
10.3389/fphy.2020.00188,
Borissova:2022mgd, 
Ruiz:2021qfp}.

The Quantum Improved Regular Kerr (QIRK) black hole model is based on the asymptotic safety framework, 
introducing a running gravitational coupling $G(r)$ that varies with energy scale
\cite{10.3389/fphy.2020.00056}. 
This approach not only resolves the Kerr black hole's ring singularity but also addresses issues such as closed timelike curves (CTCs)
\cite{carter_global_1968}, 
which would otherwise violate causality. 
Unlike classical black holes, 
which suffer from infinite curvature at singularities, 
the quantum-improved model ensures finite curvature invariants while maintaining consistent thermodynamic properties
\cite{PhysRevD.83.044041, 
chen_running_2022}. 
This is crucial, 
as it allows the black hole to obey the fundamental laws of thermodynamics even as quantum gravitational corrections stabilize its core.

The observation of light deflection in a gravitational field in 1919 provided the first experimental confirmation of a prediction from general relativity. 
This phenomenon, 
now known as gravitational lensing
\cite{Ghosh:2022mka,
AbhishekChowdhuri:2023ekr}, 
is a key method for detecting gravitational fields. 
When a photon passes near a black hole, 
it can either be absorbed,  trapped in its vicinity, or scattered to infinity. 
The observed image of such photons
is referred to as the black hole shadow
\footnote{Here, "shadow" specifically refers to the photon ring image of the black hole, not the inner shadow produced by a thin accretion disk.
}. 
The first analytic study of black hole shadows began with Synge's
\cite{Synge:1966okc} and Luminet's \cite{Luminet:1979nyg}
discussions on the Schwarzschild black hole, 
followed by Bardeen's analysis of the Kerr black hole in 1973
\cite{1973blho.conf..215B}. 
These foundational works paved the way for further exploration of black hole shadows in various models
\cite{noauthor_escape_nodate, 
grenzebach_photon_2014, 
bambi_apparent_2009,
takahashi_shapes_2004, 
wei_observing_2013, 
abdujabbarov_shadow_2013, 
atamurotov_optical_2015,
schee_optical_2009, 
tsukamoto_constraining_2014, 
takahashi_black_2005, 
bambi_direct_2012,khodadi_black_2020, 
shaikh_shadows_2019, 
falcke_toward_2013, 
tsupko_analytical_2017,
qian_cuspy_2022, 
noauthor_charged_2022, 
contreras_black_2020, perlick_calculating_2022,
falcke_viewing_1999, atamurotov_shadow_2013, vries_apparent_2000, hioki_measurement_2009,
abdujabbarov_coordinate-independent_2015, 
yumoto_shadows_2012, Cunha:2015yba,
Cunha:2018acu,
Lima:2021las,
Chen:2020aix,
2005Natur.438...62S,Cao:2024kht,
Chowdhuri:2020ipb,
Konoplya:2024lch}, 
including regular black holes
\cite{li_measuring_2014, amir_shapes_2016, kumar_shadow_2019,banerjee_signatures_2022, allahyari_magnetically_2020,
olmo_shadows_2023, abdujabbarov_shadow_2016, tsukamoto_black_2018, dymnikova_identification_2019,
noauthor_ergosphere_2020, jusufi_quasinormal_2021, ling_shadows_2022,Kumar:2018ple,Kumar:2020yem,Kumar:2020ltt,Cao:2023par,
Ghosh:2022gka,Brahma:2020eos}.

Although the black hole shadow offers a direct probe of strong gravitational fields,
observed black hole images depend not only on spacetime geometry but also on emission from accreting material and gravitational lensing.
The thin accretion disk model, with its clear physical assumptions and sensitivity to lensing effects, has become an important framework for studying black hole imaging. 
In this model, photons emitted from different regions of the disk follow geodesics to the observer,  
producing features such as a direct image of the disk, higher-order lensing rings, and the narrow photon ring.
These features encode both the black hole's spacetime geometry and the emission properties.
By studying black hole images under a thin accretion disk,  
one can separate lensing induced features such as the inner shadow (horizon image) 
and the photon ring,  
thereby measuring black hole parameters,  
and testing possible deviations from general relativity
\cite{Chael:2021rjo, Hou:2022eev, Meng:2025ivb}.
In particular, for quantum-corrected models like QIRK,  
thin-disk imaging may reveal differences from Kerr black holes,
providing a new perspective for seeking signatures of quantum gravity effects.

A major breakthrough in the study of black hole images came in 2019 
\cite{EventHorizonTelescope:2019dse, EventHorizonTelescope:2019ggy, EventHorizonTelescope:2019pgp,
EventHorizonTelescope:2019ths, EventHorizonTelescope:2019uob, EventHorizonTelescope:2019jan} 
when the Event Horizon Telescope (EHT) released the first image of the supermassive black hole M87*. 
This was followed in 2022 by EHT's observations of Sgr A*
\cite{EventHorizonTelescope:2022wkp, EventHorizonTelescope:2022xqj, EventHorizonTelescope:2022urf,
EventHorizonTelescope:2022wok, EventHorizonTelescope:2022apq,EventHorizonTelescope:2022exc}. 
Based on these EHT results, 
the images of M87* and Sgr A* have become important tools for testing and constraining theories of gravity in the strong-field regime. 
Recently, a vast body of literature has aimed at deepening our understanding of black holes through these observations
\cite{afrin_parameter_2021,
vagnozzi_hunting_2019,
afrin_estimating_2022,vagnozzi_horizon-scale_2023,
kumar_rotating_2020, kumar_gravitational_2020, walia_testing_2022, ghosh_parameters_2021,
islam_investigating_2023, noauthor_shadows_nodate, walia_observational_2023, Cunha:2019ikd,Kumar:2020yem,Kumar:2020ltt,Uniyal:2022vdu,
Pantig:2022ely,Afrin:2021wlj,Bambi:2019tjh,Shaikh:2021yux,Pal:2023wqg,
Aliyan:2024xwl,Nozari:2024jiz,Nozari:2023flq}.

Motivated by these developments,
we first simulate the images of QIRK black holes illuminated by a thin accretion disk,
to quantify how the quantum correction parameter \(\widetilde{\omega}\) affects black hole images.
Furthermore, utilizing EHT observational data for M87* and Sgr A*,
we impose constraints on the QIRK black hole parameters (spin \(a\) and quantum correction \(\widetilde{\omega}\)),
to explore their possible range of existence in the real universe.
Finally, we investigate the optical characteristics of closed timelike curves in a special, horizonless QIRK spacetime,
to better understand how quantum corrections modify the interior ($r\!\le\!0$) geometry.

This article is organized as follows:  
Section~\ref{quantum improved Regular Kerr Black Hole} establishes the framework of the QIRK,  
discussing the conditions for regularity, the existence of horizons, and the requirements to avoid closed timelike curves (CTCs) to define the physically relevant parameter space.  
Section~\ref{sec_nullgeodesics} analyses null geodesics and the photon-sphere structure.
Section~\ref{Shadow} provides a detailed analysis of QIRK black hole images illuminated by a thin accretion disk.  
Section~\ref{Constraining with EHT Observations} uses Event Horizon Telescope (EHT) observations of M87* and Sgr A* to impose constraints on the QIRK black hole parameters.  
Section~\ref{Image of Closed Timelike Curves} explores the optical features of CTCs in horizonless QIRK spacetimes.
Finally, Section~\ref{Conclusion} summarizes the main findings and 
discusses the potential of the QIRK black hole as an astrophysical black hole candidate under current and future observational capabilities.

\section{\textbf{quantum improved Regular Kerr Black Hole}}\label{quantum improved Regular Kerr Black Hole}

The asymptotically safe scenario for a quantum generalization of general relativity with a cosmological constant proposes an energy-scale \( k \) dependent Newton constant \( G(k) \) and a cosmological constant \( \Lambda(k) \). 
Assuming the cosmological constant is already at its fixed point and negligibly small, 
the solution for the Newtonian coupling derived from the associated renormalization group equations takes the form \cite{pawlowski_quantum-improved_2018}:
\begin{align}
    G(k) = \frac{G_0}{1 + \omega G_0 k^2} ,
\end{align}
where \( \omega \) is a constant of order 1 representing quantum effects, 
and \( G_0 = G(0) \) is the usual Newton constant.

Besides resolving singularities and ensuring the consistency of black hole thermodynamics, 
Ref.~\cite{chen2024quantumimprovedregularkerr} additionally shows that quantum corrections dominate near the Planck scale, motivating the identification:
\begin{align}
\label{k21}
    k^2 = \frac{\xi^2}{(r^2 + a^2)(M r)^p},
\end{align}
where \( \xi \) is a dimensionless parameter. 
By setting \( \widetilde{\omega} = \xi^2 \omega \), 
we obtain a specific form for the Newton coupling \( G(r) \), 
which allows us to construct the QIRK metric in Boyer-Lindquist coordinates
\cite{chen2024quantumimprovedregularkerr}
\begin{eqnarray} \label{metric}
    \mathrm{d}s^2 = -\left(1 - \frac{2G(r)Mr}{\Sigma} \right)\mathrm{d}t^2 + \frac{\Sigma}{\Delta }\mathrm{d}r^2 - \frac{4aG(r)Mr\sin^2\theta}{\Sigma}\mathrm{d}t\mathrm{d}\varphi + \Sigma~\mathrm{d}\theta^2 + \frac{((r^2+a^2)^2 - a^2 \Delta \sin^2\theta)\sin^2\theta}{\Sigma}\mathrm{d}\varphi^2 \,,
\end{eqnarray}
where
\begin{eqnarray}
\Sigma = r^2 + a^2 \cos^2\theta\, ,\qquad  \Delta = r^2 - 2G(r)Mr + a^2\, ,
\end{eqnarray}
and
\begin{equation}
G(r)=G_0 \frac{(Mr)^p(r^2 + a^2)}{(r^2 + a^2)(Mr)^p + \widetilde{\omega}G_0}\, .
\end{equation}
Here, \( G_0 \) is the Newton constant as before,
\( M \) is the mass parameter, 
\( a \) is the rotation parameter, 
and \( \widetilde{\omega} \) is the quantum correction parameter introduced below Eq.(\ref{k21}). 
The magnitude of \( \widetilde{\omega} \) reflects the strength of the quantum effects, 
and we require \( \widetilde{\omega} \geq 0 \). 
Obviously, the solution reduces to the case of Kerr when \( \widetilde{\omega} = 0 \).

To simplify the discussion, one can introduce three dimensionless physical quantities as follows:
\begin{align}
    x = \frac{r}{G_0 M}, \quad A = \frac{a}{G_0 M}, \quad \widetilde{\Omega} = \frac{\widetilde{\omega}}{G_0^{1+p}M^{2+2p}}.
\end{align}
With these substitutions, the function \( \Delta(r) \) becomes
\begin{align}
    \frac{\Delta(r)}{(G_0 M)^2} = x^2 - 2\frac{x^p(x^2 + A^2)}{x^p(x^2 + A^2) + \widetilde{\Omega}}x + A^2. \label{dimensionless_Delta_1}
\end{align}
In the following discussion, 
these dimensionless variables \( x \), \( A \), and \( \widetilde{\Omega} \) will be substantially used.
However, for convenience, 
we will still denote them by the original symbols \( r \), \( a \), and \( \widetilde{\omega} \).

In this section, we discuss the conditions necessary to remove singularities, guarantee the presence of horizons, 
and prevent closed timelike curves.
We then examine the implications of each condition and their physical significance. 
Finally, we integrate these results to present a comprehensive understanding of the QIRK metric.

\subsection{Resolving the Singularity and Ensuring Horizon Existence
}

As a regular black hole, 
one of the key requirements is ensuring that the spacetime remains regular, 
even at the center, 
where classical singularity would typically emerge. 
To understand how this regularity is maintained, 
we focus on resolving the singularity by examining the behavior of the metric and curvature invariants near $r=0$.

For the rotating black holes, 
the algebraically complete set consists of four invariants,
two of them, $R$ and $I_6$,
are real, while the other two, $I$ and $K$, are complex
\cite{chen2024quantumimprovedregularkerr,torres_regular_2016}.
The resolution of singularity has been discussed in detail in Ref.
\cite{chen2024quantumimprovedregularkerr}, 
thus we will only briefly explain it here. 
The behavior of the four curvature invariants
near the center of the equatorial plane from different directions,  
i.e. $r \rightarrow 0$, 
have the following fall off:
\[
R \sim \left\{
\begin{array}{ll}
r^{p-1}, & \theta \neq \frac{\pi}{2} \\
r^{p-3}, & \theta = \frac{\pi}{2}
\end{array}
\right.,
\quad
I_{6} \sim I \sim \left\{
\begin{array}{ll}
r^{2(p-1)}, & \theta \neq \frac{\pi}{2} \\
r^{2(p-3)}, & \theta = \frac{\pi}{2}
\end{array}
\right.,
\quad
K \sim \left\{
\begin{array}{ll}
r^{3(p-1)}, & \theta \neq \frac{\pi}{2} \\
r^{3(p-3)}, & \theta = \frac{\pi}{2}
\end{array}
\right..
\]

In order to avoid singularities and discontinuities, it is necessary to set $p$ as an integer greater than 3.
Furthermore, if $p$ is an odd integer, 
the term $(r^{2}+a^{2})(M r)^{p} + \omega G_{0}$, 
which appears in the denominator of all four invariants (see Eqs. (4.8) - (4.11) in Ref. \cite{chen2024quantumimprovedregularkerr}),
will have a zero point when $r<0$
(both $M$ and $\omega$ are positive).
At this zero point, these invariants will diverge,
resulting in a curvature singularity. 
Therefore, we consider the case where $p$ is an even integer greater than 3.

If we disregard the specific form of $G(r)$ and examine the behavior of the four invariants (see Eqs. (A5)-(A9) in \cite{chen2024quantumimprovedregularkerr}) as $r \rightarrow 0$, we obtain the following conditions:
\begin{eqnarray}
    G(0) = 0,~~~~~G'(0) = 0,~~~~~G''(0) = 0,~~~~~G'''(0) = 0.
\end{eqnarray}
These constraints are necessary to resolve the singularity.

Since we are interested in the QIRK black hole rather than a compact star without a horizon, 
our primary objective is to identify the parameter range within which horizons exist. 
The horizons are determined by the equation
\begin{equation}
    \Delta = r^2 - 2G(r)r + a^2 = 0 \, \label{Delta_func}.
\end{equation}

\begin{figure}[!ht]
    \centering
    \includegraphics[width = 10cm]{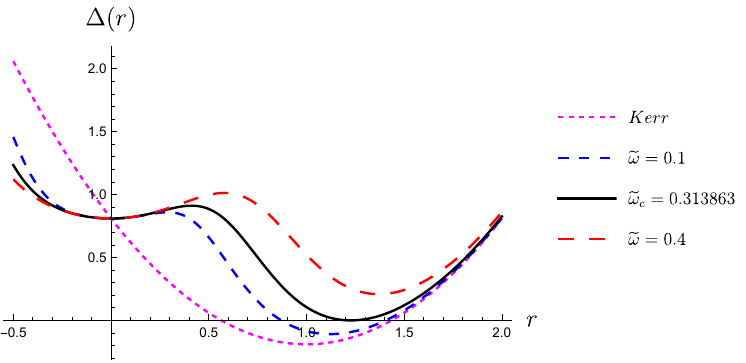}
    \caption{We set $a=0.9$ and $p=4$ to analyze the effect of $\widetilde{\omega}$ on the behavior of $\Delta(r)$.
    $\widetilde{\omega}_c$ is the critical value which two horizons will coincident.
    If $\widetilde{\omega} > \widetilde{\omega}_c$, 
    then there are no horizons.
    On the contrary, there exists at least one horizon.
    }
    \label{Delta}
\end{figure}

Fig. \ref{Delta} illustrates the relationship between $\Delta(r)$ and $r$.
As $r \rightarrow \pm \infty$, $\Delta(r)$ approaches $+\infty$, 
so only a limited range of $r$ values is shown. 
From Fig. \ref{Delta}, we observe that this black hole exhibits an additional zero point in $\Delta'(r)$ compared to the Kerr black hole with the corresponding rotation parameter $a$.

For fixed values of \( a \) and \( p \), 
Fig.~\ref{Delta} illustrates that \( \widetilde{\omega} \) reaches a critical threshold, \( \widetilde{\omega}_c \), 
at which the inner and outer horizons coincide. 
This represents the extremal case. 
If \( \widetilde{\omega} \) exceeds \( \widetilde{\omega}_c \), 
the metric describes a compact object without a horizon. 
The parameter \( \widetilde{\omega} \) reflects the strength of quantum corrections. 
Thus, 
the existence of a horizon requires moderate quantum effects, 
specifically \( \widetilde{\omega} \leq \widetilde{\omega}_c \). 
When this condition is met, 
the metric corresponds to a black hole, 
which is the primary focus here. 
It is also worth noting that Fig.~\ref{Delta} represents the particular case of \( a = 0.9 \) and \( p = 4 \). 
Because \( G(r) \) depends on the spin,
variations in \( a \) (and \( p \)) deform \( \Delta(r) \) in a non-trivial way.
Rather than analyzing the $a$-dependence case by case,
the horizon structure can be inferred from the full \( (a,\tilde{\omega}) \) plane in Fig.~\ref{range_changep} (for each fixed \( p \)).
We therefore omit a detailed discussion of the $a$-dependence here.

\subsection{Avoiding Closed Timelike Curves}

It is well known that closed timelike curves (CTCs) exist in the Kerr black hole, 
which could lead to violations of causality. 
However, the CTCs can be eliminated in this black hole model due to quantum effects. 
In this section, 
we examine the parameter constraints that arise from the avoidance of CTCs.

The issue of avoiding CTCs was also addressed in Ref.
\cite{chen2024quantumimprovedregularkerr}. 
However, 
due to the \( p \)-th power in the function \( G(r) \), 
analytical calculations become very complicated. 
Therefore, they applied certain approximations to derive a sufficient condition for avoiding CTCs:
\begin{align}
    1 < \frac{\widetilde{\omega}^{1/p}}{G_0 M^2}, \qquad \frac{a^2}{G_0} < 1.
\end{align}

Firstly, 
considering that the $g_{\varphi \varphi}$ component contains the $\Delta(r)$ term, 
even when $\theta = \pi/2$, 
its expression remains challenging to simplify due to the higher powers of $G(r)$ in $\Delta(r)$. 
Secondly, 
to accurately determine the parameter ranges that allow us to utilize the Event Horizon Telescope (EHT) data for tighter parameter constraints in Sec.~\ref{Constraining with EHT Observations}, 
we adopt a numerical approach to analyze the conditions required to avoid CTCs.

Specifically, we consider the condition $g_{\varphi \varphi} > 0$ to avoid CTCs:
\begin{align}
    g_{\varphi \varphi} = \frac{((r^2 + a^2)^2 - a^2 \Delta \sin^2\theta) \sin^2\theta}{\Sigma}. \label{gphiphi}
\end{align}

A sufficient condition can be obtained by setting $\theta = \pi/2$. Since $G(r)$ is always positive, 
we only need to consider the case where $r < 0$. 
In this case, Eq.~(\ref{gphiphi}) reduces to:
\begin{align}
    g_{\varphi \varphi} = \frac{r^3 + a^2 r + 2 a^2 G(r)}{r} \sim \left\{
        \begin{array}{ll}
            +\infty, & r \rightarrow -\infty, \\
            a^2, & r \rightarrow 0.
        \end{array}
    \right.
\end{align}

\begin{figure}[!ht]
    \centering
    \includegraphics[width=10cm]{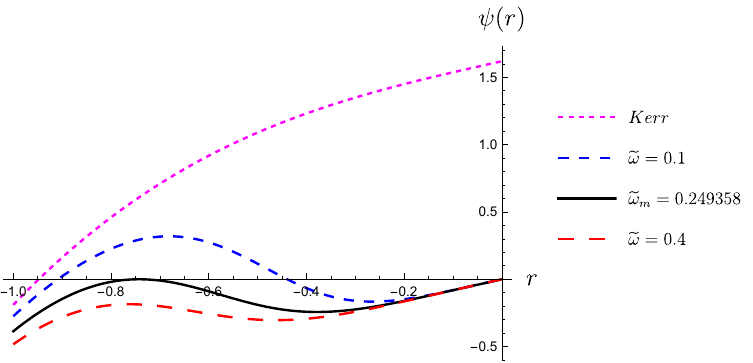}
    \caption{Plot of $\psi(r)$ as a function of $r$ for $a = 0.9$ and $p = 4$. When $\widetilde{\omega} > \widetilde{\omega}_m$, $\psi(r)$ has no real zeros, corresponding to $g_{\varphi \varphi} > 0$ for all $r$. Conversely, if $\psi(r)$ has a zero point, there will exist some $r$ for which $g_{\varphi \varphi} < 0$, implying the existence of CTCs.}
    \label{psir}
\end{figure}

Next, let us consider the polynomial $\psi(r) = r^3 + a^2 r + 2 a^2 G(r)$. 
To ensure $g_{\varphi \varphi} > 0$ for $r < 0$, $\psi(r)$ must have no zero points. 
Fig.~\ref{psir} plots $\psi(r)$ for $a = 0.9$ and $p = 4$ with varying $\widetilde{\omega}$ values. 
As shown, a minimum value of $\widetilde{\omega}$, denoted $\widetilde{\omega}_m$, 
exists such that $\psi(r)$ is tangent to the $r$-axis. 
Thus, to avoid CTCs, $\widetilde{\omega}$ must satisfy $\widetilde{\omega} > \widetilde{\omega}_m$. 

Therefore, 
in order to avoid CTCs, 
a lower bound on the quantum correction was proposed, 
specifically requiring \( \widetilde{\omega} > \widetilde{\omega}_m \).
Additionally, $\widetilde{\omega}_m$ exists for other values of $a$ and $p$, similar to the previous discussion regarding $\widetilde{\omega}_c$.

\begin{figure}[!ht]
    \centering
    \begin{minipage}{6.35cm}
        \centering
        \includegraphics[width=\textwidth]{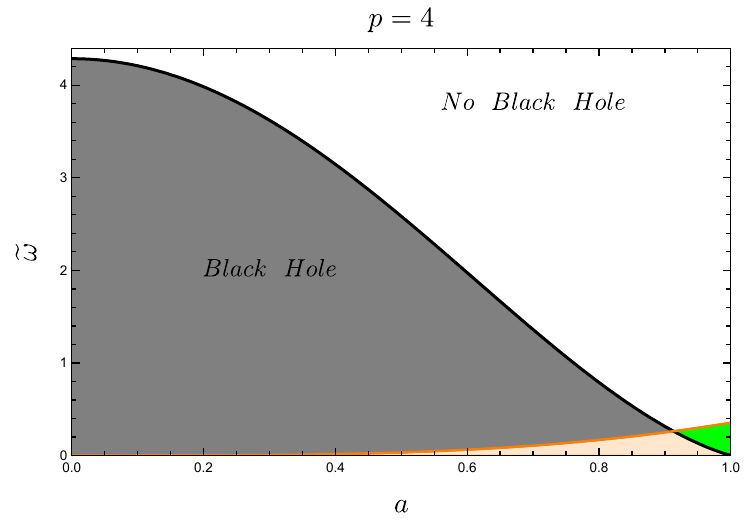}
    \end{minipage}
    \hfill 
    \begin{minipage}{11.44cm}
        \centering
        \includegraphics[width=\textwidth]{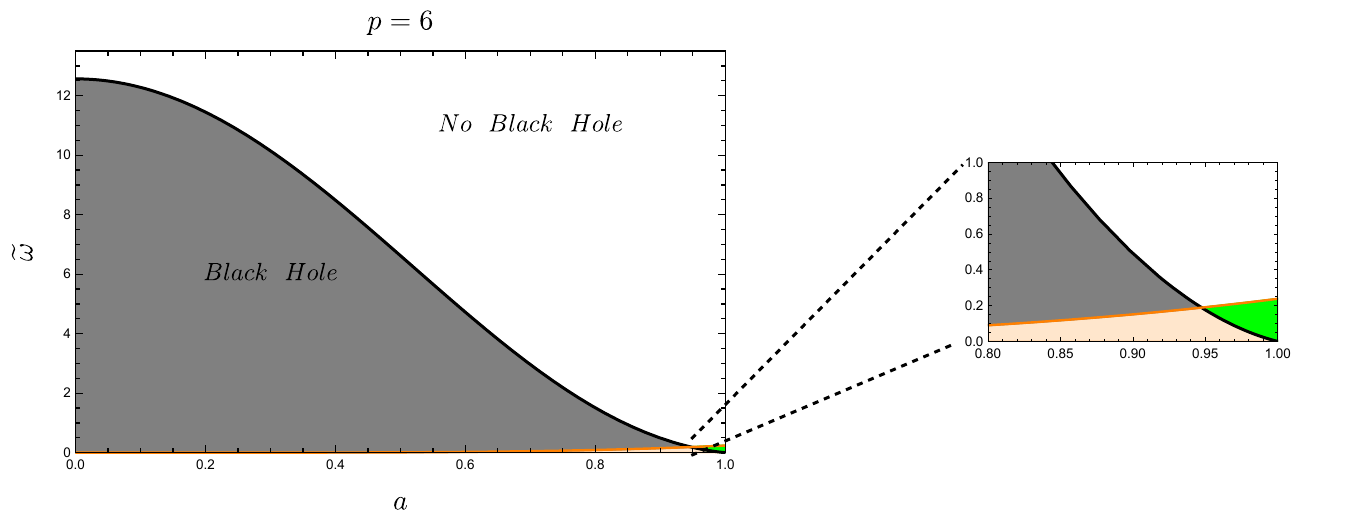}
    \end{minipage}
    \caption{Parameter plane $(a, \widetilde{\omega})$ for the QIRK spacetime. 
    The black curve corresponds to $\widetilde{\omega}_c$, which separates the black hole spacetimes from the horizonless spacetimes. 
    The orange curve represents $\widetilde{\omega}_m$, which divides spacetimes with closed timelike curves (CTCs) from those without CTCs. 
    The grey region and white region represent black holes and horizonless compact objects without CTCs, respectively. 
    The orange region and green region represent black holes and horizonless compact objects with CTCs, respectively.}
    \label{range_changep}
\end{figure}

While ensuring that the above three conditions are satisfied, 
we obtain the parameter range illustrated in Fig.~\ref{range_changep}. 
The black curve represents \( \widetilde{\omega} = \widetilde{\omega}_c \), 
where the horizons coincide, 
and the orange curve corresponds to \( \widetilde{\omega} = \widetilde{\omega}_m \), 
where \( \psi(r) \) becomes tangent to the \( r \)-axis. 
The region of interest is where \( \widetilde{\omega} \) lies between \( \widetilde{\omega}_c \) and \( \widetilde{\omega}_m \), 
as indicated by the shaded region in Fig.~\ref{range_changep}. 

Additionally, the green-shaded area denotes the region where no horizon exists, but CTCs are present.  
This suggests that certain causality-violating events or signals may be observed by external observers.  
While such a region is allowed by the QIRK metric, we discuss images of these closed timelike curves in the final part of this paper.

It is also important to note that the black and orange curves intersect at a rotation parameter \( a < 1 \). 
This implies that, for the QIRK black hole, 
there exists a maximum rotation parameter \( a_{\text{max}} \), 
which is strictly less than 1. 
This behavior is different from the Kerr black hole, where extremality is achieved at \( a = 1 \). 
The specific values of \( a_{\text{max}} \) are provided in Table~\ref{table}.

In this paper, 
we focus on the cases of \( p = 4 \) and \( p = 6 \) for the following reasons: 
first, as illustrated in Fig.~\ref{range_changep}, 
as the value of \( p \) increases, 
the parameter range for the QIRK black hole expands, 
and the orange curve representing the avoidance of CTCs approaches the \( a \)-axis, 
with \( a_{\text{max}} \) gradually approaching 1. 
This could potentially obscure certain distinctive features of the QIRK black hole. 
Second, from the aspect of black hole shadow
(not shown here),
a smaller value of \( p \) results in greater deviations from the Kerr black hole. 
Thus, considering smaller values of \( p \) better highlights the differences between the QIRK black hole and the Kerr black hole.

\begin{table}[]  
    \centering  
    \caption{Maximum rotation parameter \(a_{\text{max}}\) for different values of \(p\).
    Here \(a_{\text{max}}\) refers to the spin at which the \(\widetilde{\omega}_c\) and \(\widetilde{\omega}_m\) curves intersect, as shown in Fig.~\ref{range_changep}.
    } 
    \setlength{\arraycolsep}{10pt} 
    \begin{tabular}{c@{\hspace{2 em}}c@{\hspace{2 em}}c@{\hspace{2 em}}c@{\hspace{2 em}}c@{\hspace{2 em}}c@{\hspace{2 em}}c@{\hspace{2 em}}c@{\hspace{2 em}}c@{\hspace{2 em}}c} 
    \hline
    \hline  
    p                           & 4             & 6             &   8           & 10          & 12    &14           &16          \\ \hline  
    $a_{\text{max}}$ & 0.913141       & 0.947498      &   0.964022   &0.973442    & 0.979401 &0.983448    &0.986342           \\ \hline  
    \end{tabular}  
    \label{table}
\end{table}

\section{Null geodesics}\label{sec_nullgeodesics}

Photons follow null geodesics in QIRK spacetime,  
and these paths govern lensing and the formation of black hole images.  
By solving for unstable photon orbits and critical impact parameters,  
we quantify how quantum corrections alter light deflection relative to Kerr.  
This is essential for linking the QIRK geometry to observable imaging features.

Before proceeding with the discussion, we specify the spacetimes under consideration.
Within stationary, axisymmetric separable geometries (for which photon geodesics separate), there exists a subclass whose line element is Kerr-like and differs only by a modified radial function $\Delta(r)$;
the QIRK spacetime belongs to this subclass.
We have already characterized $\Delta$ for QIRK in Fig.~\ref{Delta}.
Since the analysis below is independent of its explicit form, we do not substitute it here—doing so would make the formulas needlessly long and less transparent.
We emphasize, however, that this Kerr-like subclass represents only a subset of Liouville-integrable rotating solutions,
many other integrable families also admit separable geodesics but exhibit qualitatively different radial and polar potentials.


Photon trajectories within the spacetime described by the metric (\ref{metric}) are governed by the geodesic equations, 
which can be derived from the Hamilton-Jacobi equation \cite{1968PhRv..174.1559C}:
\begin{align}
    \frac{\partial S}{\partial \lambda} = -\frac{1}{2} g^{\alpha \beta} \frac{\partial S}{\partial x^\alpha} \frac{\partial S}{\partial x^\beta},
\end{align}
where \( \tau \) is the affine parameter along the geodesics, and \( S \) is the Jacobi master function. 
The metric (\ref{metric}) exhibits both time translational and rotational invariance, 
which leads to conserved quantities along the geodesics: 
the energy \( E = -p_t \) and the axial angular momentum \( L = p_\varphi \), 
where \( p_{\mu} \) represents the photon's four-momentum. 
The Petrov-type D nature of the metric (\ref{metric}) guarantees the existence of Carter's separable constant \( \mathcal{K} \), 
allowing the action to be written in the form:
\begin{align}
    S = -E t + L \varphi + S_r(r) + S_{\theta}(\theta),
\end{align}
where \( S_r(r) \) and \( S_{\theta}(\theta) \) are functions only of \( r \) and \( \theta \), respectively. 
The geodesic equations can then be written as the following complete set of first-order differential equations for null geodesics
\cite{1968PhRv..174.1559C,Chandrasekhar_1985}:
\begin{align}
    \Sigma \frac{\mathrm{d}t}{\mathrm{d}\lambda } &=a(L-a E\sin^2\theta )+\frac{r^2+a^2}{\Delta }\Big[(r^2+a^2)E-a L\Big]\, ,\\
    \Sigma \frac{\mathrm{d}\varphi }{\mathrm{d}\lambda } &=\frac{L}{\sin^2\theta } -a E +\frac{a}{\Delta }\Big[(r^2+a^2)E-a L\Big]\, ,\\
    \Sigma \frac{\mathrm{d}r }{\mathrm{d}\lambda } &= \pm \sqrt{R(r)}\, , \label{r_eq}\\
    \Sigma \frac{\mathrm{d}\theta  }{\mathrm{d}\lambda } &= \pm \sqrt{\Theta (\theta )}\, , \label{theta_eq}
\end{align}
where
\begin{align}
    R(r) &=\Big[(r^2+a^2)E -a L\Big]^2 -\Delta \Big[\mathcal{K} +(L-a E)^2\Big]\, ,\\
    \Theta (\theta ) &= \mathcal{K}+\Big(a^2E^2 -\frac{L^2}{\sin^2\theta }\Big)\cos^2\theta\, . \label{Theta_motion}
\end{align}
The "+" and "-" signs in Eq.~(\ref{r_eq}) correspond to outgoing and ingoing photons in the radial direction, 
while in Eq.~(\ref{theta_eq}) they represent photons moving towards the south pole \( \theta = \pi \) and the north pole \( \theta = 0 \), respectively. 
The geodesic equations described above govern the propagation of photons in the QIRK spacetime.
When studying photon orbits, 
two distinct impact parameters are typically defined
\cite{Chandrasekhar_1985}:
\begin{align}\label{xieta_def}
    \xi = \frac{L}{E}, \quad \eta = \frac{\mathcal{K}}{E^2} .
\end{align}
These two impact parameters play a crucial role in determining the shape of a photon's orbit. 
Based on an analysis of the radial effective potential $\mathcal{V}_r(r)$, 
photon orbits can be classified into three distinct categories: 
scattered orbits, absorbed orbits, and bounded orbits (also known as spherical orbits or photon sphere). 
If a spherical orbit is unstable, small perturbations allow photons to escape and create the observable photon ring image.
These orbits are determined by the following conditions:
\begin{eqnarray}\label{unstable_condition}
    R(r)=0\, ,\quad\quad\frac{\mathrm{d}R(r)}{\mathrm{d}r}=0\, ,
\end{eqnarray}
which can be written exactly with impact parameters $\xi$ and $\eta$
\begin{align}  
    \frac{R(r)}{E^2} &= \left(a^2-a \xi +r^2\right)^2-\Delta (r) \left[ (\xi -a)^2+\eta \right] = 0, \label{zero derivative of R} \\  
    \frac{R'(r)}{E^2} &= 4 r \left(a^2-a \xi +r^2\right)-\left[ (a-\xi )^2+\eta \right] \Delta '(r) = 0. \label{first derivative of R}  
\end{align} 
When we consider a nonzero rotation parameter,
the solutions of (\ref{zero derivative of R}) and (\ref{first derivative of R}) include an additional
family compared to the Kerr case.
One such family is  
\begin{align}
    \xi &= \frac{a^2 \Delta '(r)+r^2 \Delta '(r)-4 r \Delta (r)}{a \Delta '(r)}\Big|_{r = r_{\text{sphere}}}\, ,  \label{xi_sol} \\
    \eta &= \frac{r^2 \left(16 a^2 \Delta (r)-r^2 \Delta '(r)^2+8 r \Delta (r) \Delta '(r)-16 \Delta (r)^2\right)}{a^2 \Delta '(r)^2}\Big|_{r = r_{\text{sphere}}}\, . \label{eta_sol}
\end{align}
The first family coincides with the Kerr solution; the second is specified by \( r = 0\) and \( \eta = 0\).
At the same time, 
it is not hard to find
\begin{eqnarray} 
    R''(r_{\text{sphere}}) &=& 4 \left(a^2-a \xi +3 r^2\right)-\left(a^2-2 a \xi +\eta +\xi ^2\right) \Delta ''(r)\Big|_{r = r_{\text{sphere}}}\, \\
                    &=& 8 r \left(\frac{2 \Delta (r) \left(\Delta '(r)-r \Delta ''(r)\right)}{\Delta '(r)^2}+r\right)\Big|_{r = r_{\text{sphere}}} \, . \label{second derivative of R}
\end{eqnarray}
The photon sphere is unstable when $R''(r_{\text{sphere}}) > 0$ and stable when $R''(r_{\text{sphere}}) < 0$.
For the family with \(r = 0,\ \eta = 0\), stability depends on \(\xi\): it is unstable when \(\lvert \xi \rvert < a\) and stable when \(\lvert \xi \rvert > a\).

In the case of a non-rotating black hole ($a = 0$), 
the system exhibits spherical symmetry, 
allowing us to set $\theta = \pi/2$. 
In this case, 
the motion in the $\theta$-direction reduces to $\mathcal{K} = 0$ (equivalently $\eta = 0$). 
Meanwhile,
the impact parameter $\xi$ corresponds to $b$, 
which is the impact parameter in the spherical symmetric case. 
Consequently, 
the equations $R(r) = 0$ and $R'(r) = 0$ reduce to:
\begin{align}
    4 \Delta(r) - r \Delta'(r) = 0, \\
    \xi^2 = \frac{r^4}{\Delta(r)} \Big|_{r = r_{\text{sphere}}}.
\end{align}
The \(\theta\)-equation is identical to that in Kerr.
It has been already discussed in great detail in Ref.~\cite{Gralla:2019ceu},
and we directly adopt the results here.
\begin{table}[h!]
\centering
\caption{Classification of photon trajectories based on the value of \( r_{\max} \).
    Here, \(r_+\) denotes the outer event horizon of the black hole.  
    }
\begin{tabular}{|c|l|l|}
\hline
\text{Region} & \text{Black Hole Case} & \text{Compact Object Case} \\
\hline
I   & \(r_{\max} > r_+\) & \(r_{\max} > 0\) \\
II  & \(0 < r_{\max} < r_+\) & \(r_{\max} > 0\) \\
III & \(0 < r_{\max} < r_+\) & \(r_{\max} > 0\) \\
IV  & \(r_{\max} < 0\) & \(r_{\max} < 0\) \\
V   & \(r_{\max}\) is complex, i.e., light travels from \(-\infty\) to \(+\infty\) & Same as black hole case \\
\hline
\end{tabular}
\label{rmax_regions}
\end{table}

\begin{figure}[htbp]
    \begin{minipage}{0.45 \textwidth}
        \centering
        \includegraphics[width=\linewidth]{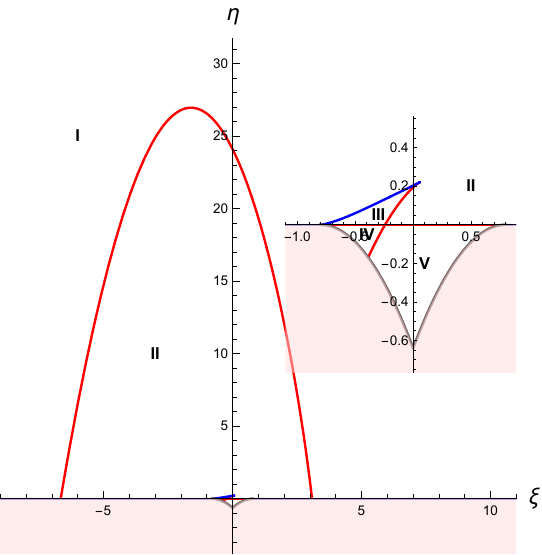}
    \end{minipage}
    \hfill
    \begin{minipage}{0.45 \textwidth}
        \centering
        \includegraphics[width=\linewidth]{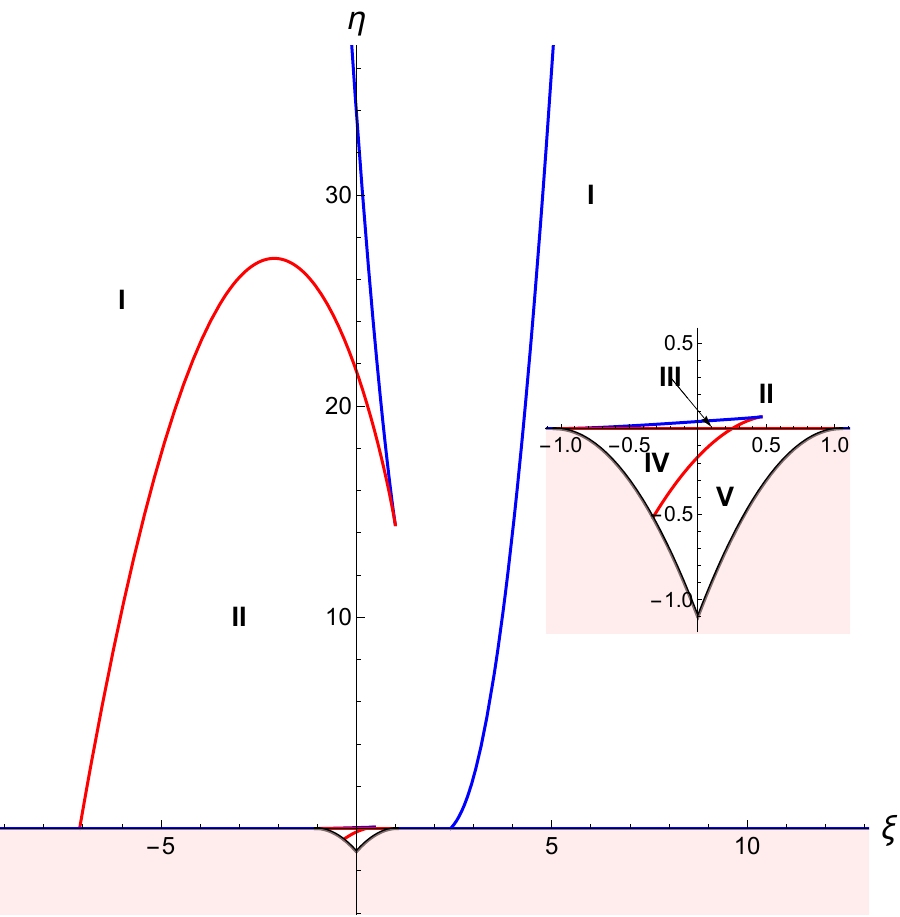}
    \end{minipage}
    \caption{The \(\xi,\eta\) parameter plot for a black hole (\(a=0.8,p=4,\widetilde{\omega}=0.7\), left) and a compact object (\(a=1.05,p=4,\widetilde{\omega}=0.1\), right).
    Here \( \xi \) and \( \eta\) are the usual impact parameters (see Eq.~\ref{xieta_def}).
    The red curves represent unstable photon spheres.
    The blue curves represent stable photon spheres.
    Compared with Kerr, 
    QIRK admits an extra photon sphere at \(\eta=0,r=0\).
    In this case, \(\lvert \xi \rvert < a\) is unstable, \(\lvert \xi \rvert > a\) is stable
    }
\label{lambdaeta}
\end{figure}

Fig.~\ref{lambdaeta} shows the \((\xi,\eta)\) parameter plot for photon trajectories in both the black hole and compact object cases.  
Red curves denote unstable photon spheres, while blue curves denote stable photon spheres.  
These curves divide the parameter space into five regions, classified by the largest real root \(r_{\max}\) of the radial potential \(R(r)\).  
Since we focus on light received or emitted by an external observer at large \(r\),  
only these regions are relevant, and the detailed classification is presented in Table~\ref{rmax_regions}.  
The red-shaded region, as in the Kerr case, is excluded by the \(\theta\)-motion.  
As stated earlier,
QIRK introduces an additional family of photon spheres compared to the Kerr case
\cite{Gralla:2019ceu},
which appears as the straight line \(\eta=0\) in the figure,
and is unstable when \(\lvert \xi \rvert < a\),
and stable when \(\lvert \xi \rvert > a\).

\section{QIRK Black hole image with a thin accretion disk}\label{Shadow}

Observations of black hole images offer a direct way to probe extreme gravitational fields,  
test gravitational theories,  
and reveal the fundamental properties of black holes. 
However,
interpreting these images requires separating the effects of spacetime geometry from the complex radiation of the surrounding accreting matter.  
In this context,  
modeling black hole images with a thin accretion disk is useful because
it simplifies the distribution and emission properties of the disk,  
the thin-disk model highlights fine features caused by strong gravitational lensing,  
such as the inner shadow and photon ring.  
The QIRK black hole,  
constructed within the asymptotic safety framework,  
removes the ring singularity of the classical Kerr solution and introduces quantum corrections that alter the spacetime geometry.  
Consequently,  
the detailed shape, size, and location of the photon ring in a QIRK image illuminated by a thin disk directly encode information about these quantum corrections.  
This provides a basis for testing quantum gravity models,  
distinguishing QIRK black holes from Kerr black holes,  
and constraining quantum gravity parameters through observation.  

Since the setup and methods have been detailed in previous works \cite{Hou:2022eev, Meng:2025ivb},  
we provide only a brief summary here.  
We place a distant zero-angular-momentum-observer (ZAMO) at (\(t_{o} = 0\), \(r_{o}\), \(\theta_{o}\), \(\phi_{o} = 0\))
and define the local orthonormal tetrad \(e_{(\mu)}\) as  
\begin{align}
e_{(0)} &= \sqrt{\frac{-\,g_{\phi\phi}}{g_{tt}\,g_{\phi\phi} - g_{t\phi}^{2}}}\,\biggl(1,\,0,\,0,\,-\tfrac{g_{t\phi}}{g_{\phi\phi}}\biggr),  
&  
e_{(1)} &= \bigl(0,\,-\tfrac{1}{\sqrt{g_{rr}}},\,0,\,0\bigr), \\[0.5ex]
e_{(2)} &= \bigl(0,\,0,\,\tfrac{1}{\sqrt{g_{\theta\theta}}},\,0\bigr),  
&  
e_{(3)} &= \bigl(0,\,0,\,0,\,-\tfrac{1}{\sqrt{g_{\phi\phi}}}\bigr).
\end{align}
The photon's four-momentum $k_{\mu}$ can be projected onto this tetrad as
\begin{equation}\label{PtoRealP}
p_{(\mu)} = k_{\nu}\,e^{\nu}{}_{(\mu)},  
\quad  
p^{(\mu)} = \eta^{(\mu)(\sigma)}\,e^{\nu}{}_{(\sigma)}\,k_{\nu}\, ,
\end{equation}
then we can define the so-called celestial coordinates \(\Theta\) and \(\Psi\) in the ZAMO frame to label each light ray, i.e., 
\begin{align}\label{ThetaPhiToP}
\cos\Theta &= \frac{p^{(1)}}{p^{(0)}},  
\quad  
\tan\Psi = \frac{p^{(3)}}{p^{(2)}}.
\end{align}
Using a stereographic projection, one can map these angles onto a two-dimensional screen via  
\begin{equation}\label{xyToThetaPhi}
x = -\,2\,\tan\frac{\Theta}{2}\,\sin\Psi,  
\quad  
y = -\,2\,\tan\frac{\Theta}{2}\,\cos\Psi.
\end{equation}
Given the screen coordinates \(x\) and \(y\),  
we first compute the photon's four momentum using (\ref{PtoRealP}), (\ref{ThetaPhiToP}) and  (\ref{xyToThetaPhi}).  
Then, by ray tracing, we obtain the photon's trajectory,  
from which we extract the redshift and intensity information carried by the photon.

We model the illumination source as an optically and geometrically thin accretion disk in the equatorial plane (\(\theta = \pi/2\)).  
Tracing light rays backward from the observer,  
they can intersect the equatorial accretion disk multiple times.  
At each intersection, the ray collects additional emission and contributes to the detected intensity.
Thus, it is essential to identify the radius \(r_{n}(x,y)\) of the \(n\)-th intersection during ray tracing. 
In particular, \(n = 1\) produces the direct emission image,  
\(n = 2\) yields the lensed ring image,  
and \(n \ge 3\) corresponds to the photon ring image.

\begin{figure}[!ht]
    \centering
    \includegraphics[width = 0.45 \linewidth]{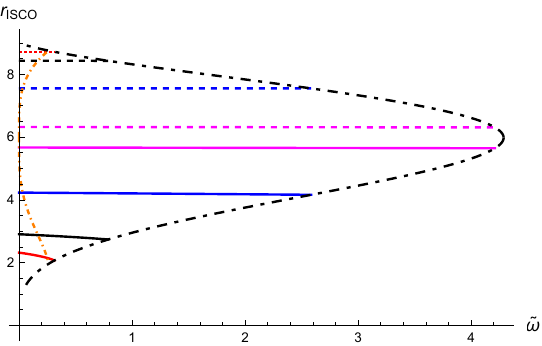}
    \caption{The radius of the ISCO as a function of the quantum correction parameter \(\widetilde{\omega}\).  
    Dashed curves represent retrograde orbits, while solid curves represent prograde orbits.  
    The magenta, blue, black, and red curves correspond to spin parameters \(a=0.1\), \(0.5\), \(0.8\), and \(0.9\), respectively.  
    Additionally, the black dash-dotted curves mark the ISCO radius at \(\widetilde{\omega} = \widetilde{\omega}_c(a)\) for each spin (the extremal black hole), and the orange dash-dotted curves mark the ISCO radius at \(\widetilde{\omega} = \widetilde{\omega}_m(a)\) (the threshold for avoiding closed timelike curves).    
    }
    \label{ISCOPlot}
\end{figure}

The observed intensity \(I_{\text{obs}}\) at a frequency \(\nu_0\) is obtained  by summing the contributions from each intersection of the light ray with the disk. 
Each contribution is the disk-emitted specific intensity \(I_{\mathrm{em}}\) multiplied by the redshift factor \(g_{n} = \nu_{\mathrm{o}} / \nu_{em}\)
\cite{Hou:2022eev}:
\begin{equation}
    I_{\text{obs}}(\nu_0, \alpha, \beta) = \sum_{n=1}^{N_{\text{max}}} g_n^3 I_{\text{em}}(\nu_{\text{em}}, r_n),
\end{equation}
where $I_{\text{em}}(r_n)$ represents the specific intensity emitted by the accretion disk material in its rest frame at radius $r_n$. 
In this work, we adopt a phenomenological emissivity model commonly used in black hole imaging simulations 
\cite{Chael:2021rjo}.
Specifically, the emissivity is taken to follow a simple radially decreasing profile,
and is independent of the emitted frequency over the narrow band of observation for simplicity.
A common functional form for such a profile, 
as used to model M87* and Sgr A*, 
involves an exponential decay with a logarithmic radial dependence
\cite{Chael:2021rjo, Hou:2022eev}
\begin{align}\label{emmisionprofile}
I_{em}(r) = \exp\left[ -\frac{1}{2} \left(\log \frac{r}{r_h}\right)^2 - 2 \left(\log \frac{r}{r_h}\right) \right].
\end{align}
The redshift factor \(g_n\) depends on the photon's four-momentum \( k_\nu\) and the four velocity  of the emitting material in the disk \(u^\mu_{\text{disk}}\):
\begin{equation}
    g_n = \frac{k_\mu u^\mu_{\text{obs}}}{k_\nu u^\nu_{\text{disk}}},
\end{equation}
where \(u^\mu_{\text{obs}}\) is the observer's four velocity. 
To determine the four-velocity of the emitting material $u^\mu_{\text{disk}}$, 
we adopt a standard model for the accretion flow model. 
Material outside the innermost stable circular orbit (ISCO) follows stable circular orbits in the equatorial plane. 
Upon reaching the ISCO,  
the material plunges inward toward the event horizon,  
following geodesics that preserve its energy and angular momentum at the ISCO. 
Fig.~\ref{ISCOPlot} shows the radius of the ISCO as a function of the quantum correction parameter \(\widetilde{\omega}\).
As \(\widetilde{\omega}\) increases,
the ISCO radius shrinks for both prograde and retrograde orbits.
However,
these changes are small,
with only the prograde orbits at high spin values exhibiting relatively noticeable shifts.

The effects of observer inclination \(\theta_{o}\) and black hole spin \( a \) have been studied previously in
\cite{Hou:2022eev},
following the discussion there,
we place the observer at a distance \(r_{o} = 100\),  
with inclination \(\theta_{o} = 80^\circ\),  
and investigate the impact of quantum corrections on the image at different spin values.
Moreover, because the parameter \(p\) has only a minor effect and larger \(p\) complicates the analysis,  
we set \(p = 4\) and focus on varying the quantum correction parameter \(\widetilde{\omega}\).

\begin{figure}[htbp]
    \begin{minipage}{0.22 \textwidth}
        \centering
        \includegraphics[width=\linewidth]{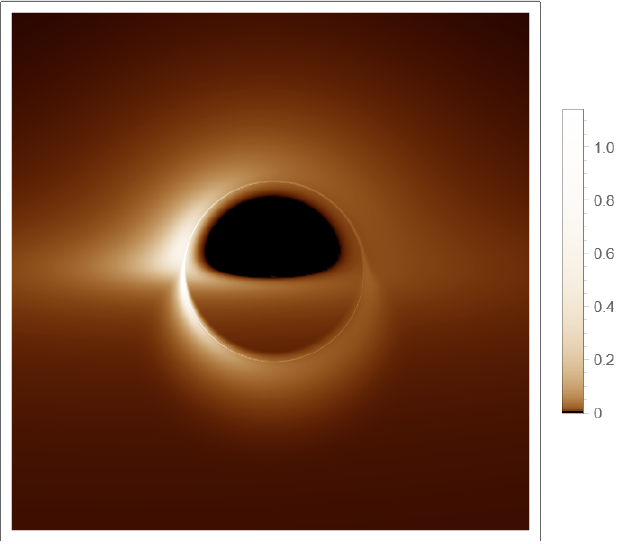}
        \vspace{0.5em}
        \tiny{(a) Kerr $a = 0.1$}
    \end{minipage}
    \hfill
    \begin{minipage}{0.22 \textwidth}
        \centering
        \includegraphics[width=\linewidth]{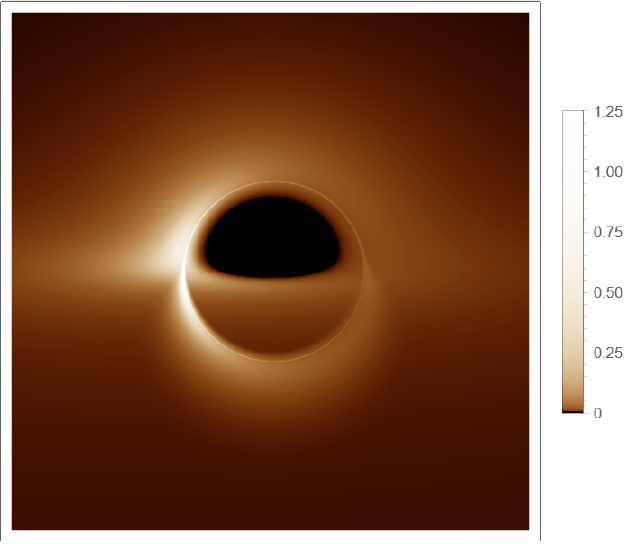}
        \vspace{0.5em}
        \tiny{(b) $a = 0.1$ $p = 4$ $\widetilde{\omega} = 0.7$}
    \end{minipage}
    \hfill
    \begin{minipage}{0.22 \textwidth}
        \centering
        \includegraphics[width=\linewidth]{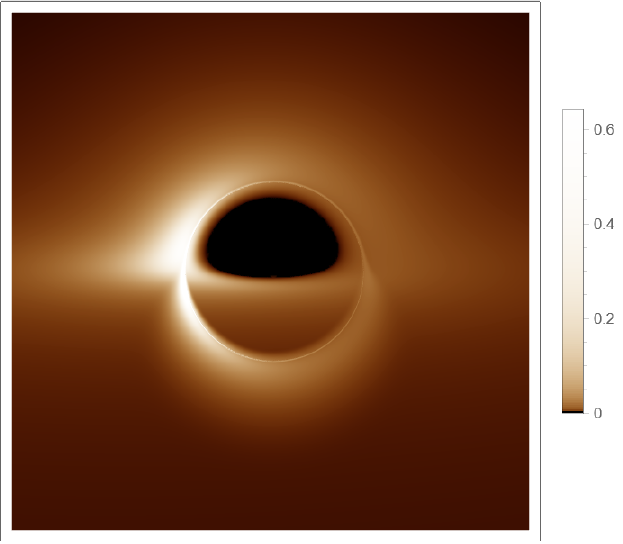}
        \vspace{0.5em}
        \tiny{(c) $a = 0.1$ $p = 4$ $\widetilde{\omega} = 1.9$}
    \end{minipage}
    \hfill
    \begin{minipage}{0.22 \textwidth}
        \centering
        \includegraphics[width=\linewidth]{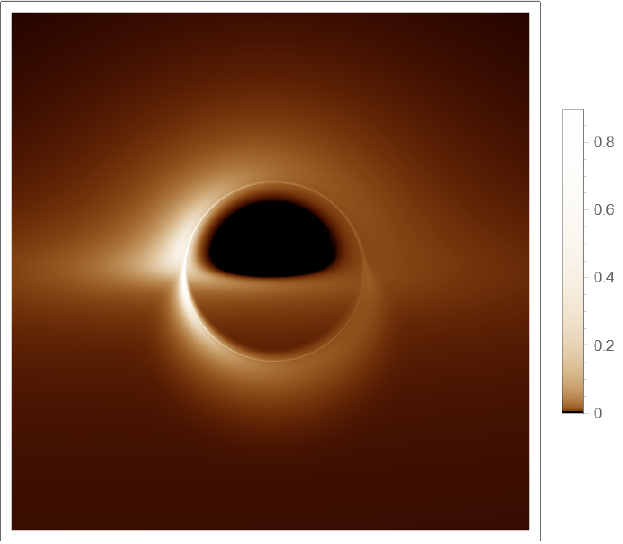}
        \vspace{0.5em}
        \tiny{(d) $a = 0.1$ $p = 4$ $\widetilde{\omega} = 4$}
    \end{minipage}
    \hfill
    \begin{minipage}{0.22  \textwidth}
        \centering
        \includegraphics[width=\linewidth]{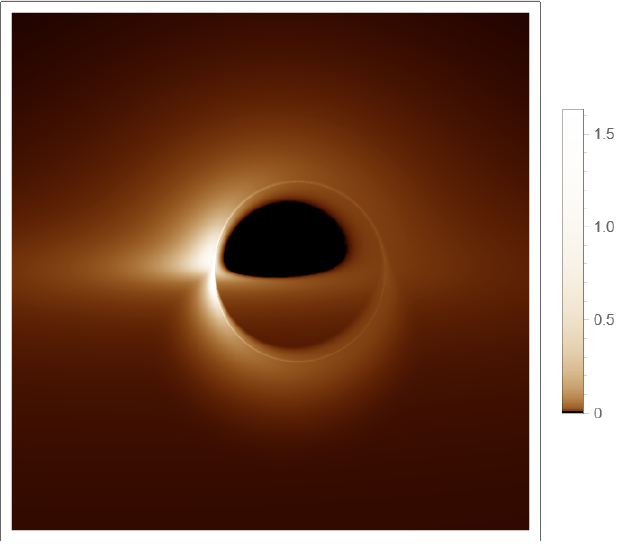}
        \vspace{0.5em}
        \tiny{(e) Kerr $a = 0.8$}
    \end{minipage}
    \hfill
    \begin{minipage}{0.22 \textwidth}
        \centering
        \includegraphics[width=\linewidth]{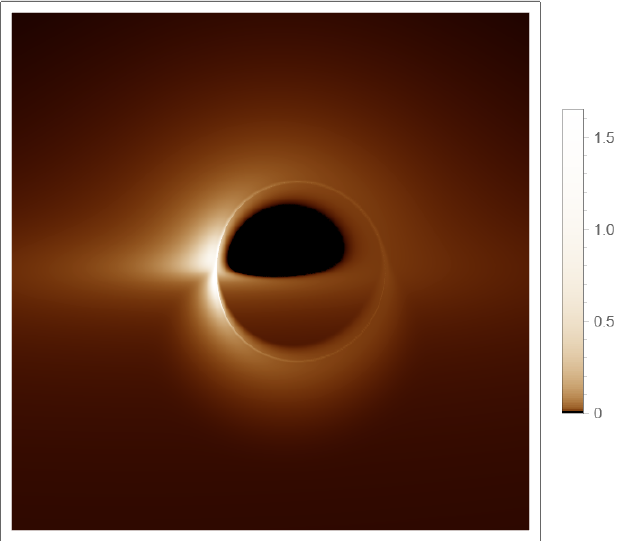}
        \vspace{0.5em}
        \tiny{(f) $a = 0.8$ $p = 4$ $\widetilde{\omega} = 0.7$}
    \end{minipage}
    \hfill
    \begin{minipage}{0.22 \textwidth}
        \centering
        \includegraphics[width=\linewidth]{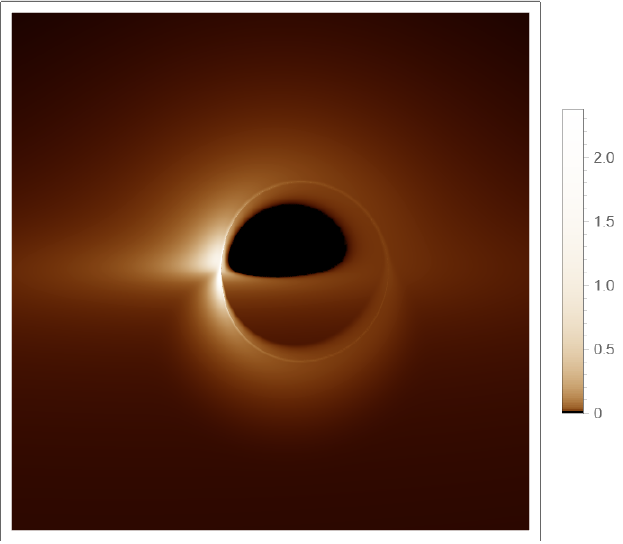}
        \vspace{0.5em}
        \tiny{(g) Kerr $a = 0.9$}
    \end{minipage}
    \hfill
    \begin{minipage}{0.22 \textwidth}
        \centering
        \includegraphics[width=\linewidth]{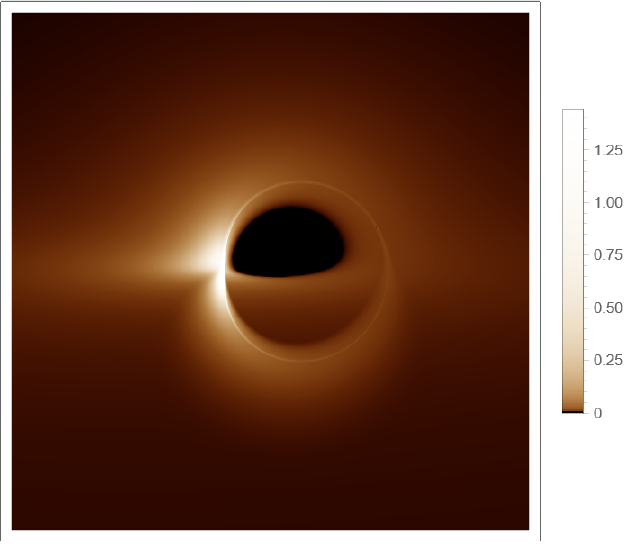}
        \vspace{0.5em}
        \tiny{(h) $a = 0.9$ $p = 4$ $\widetilde{\omega} = 0.3$}
    \end{minipage}
    \caption{ Images of the QIRK black hole illuminated by prograde accretion flows for selected model parameters.}
\label{intensityplot}
\end{figure}

\begin{figure}[htbp]
    \begin{minipage}{0.32 \textwidth}
        \centering
        \includegraphics[width=\linewidth]{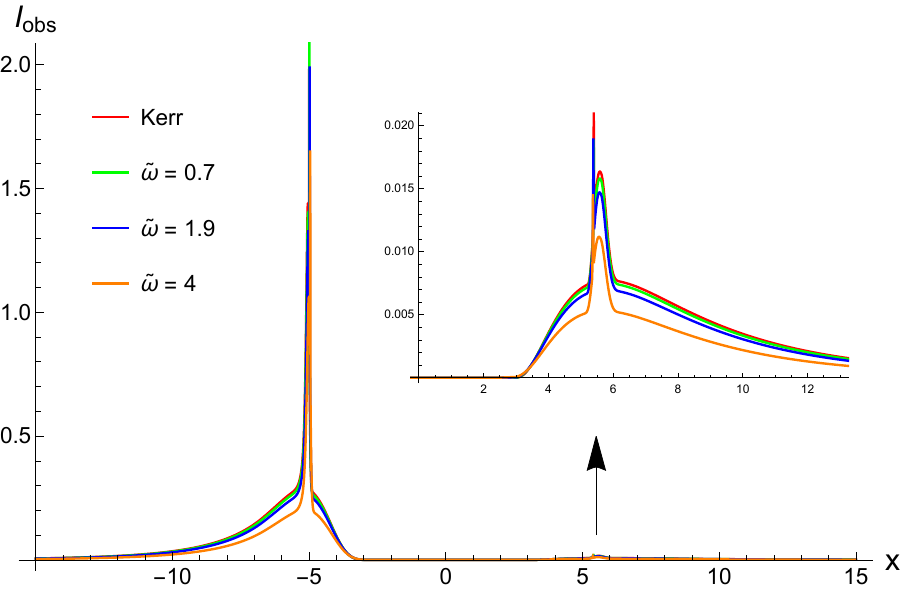}
        \vspace{0.5em}
        \tiny{(a) $a = 0.1$}
    \end{minipage}
    \hfill
    \begin{minipage}{0.32 \textwidth}
        \centering
        \includegraphics[width=\linewidth]{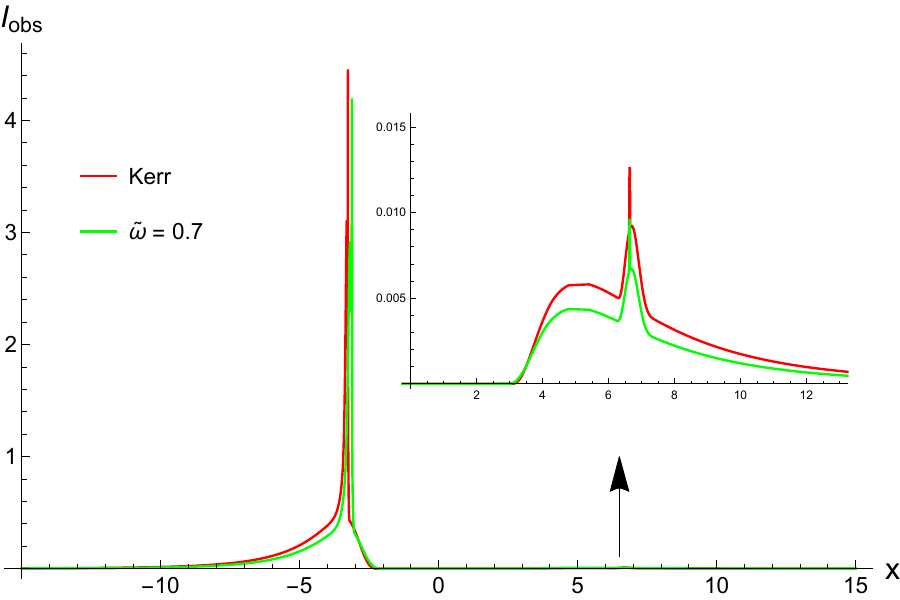}
        \vspace{0.5em}
        \tiny{(b) $a = 0.8$}
    \end{minipage}
    \hfill
    \begin{minipage}{0.32 \textwidth}
        \centering
        \includegraphics[width=\linewidth]{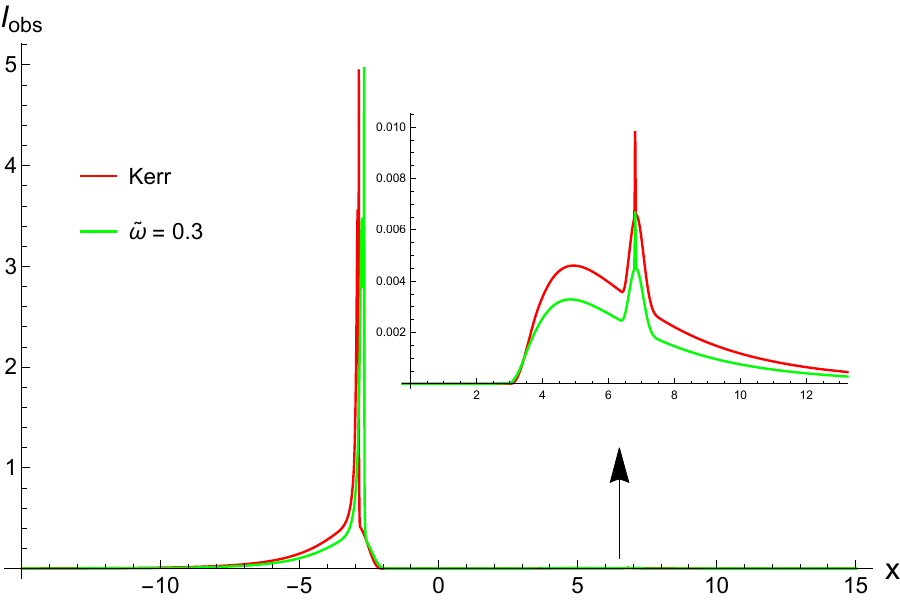}
        \vspace{0.5em}
        \tiny{(c) $a = 0.9$}
    \end{minipage}
    \hfill
    \begin{minipage}{0.32 \textwidth}
        \centering
        \includegraphics[width=\linewidth]{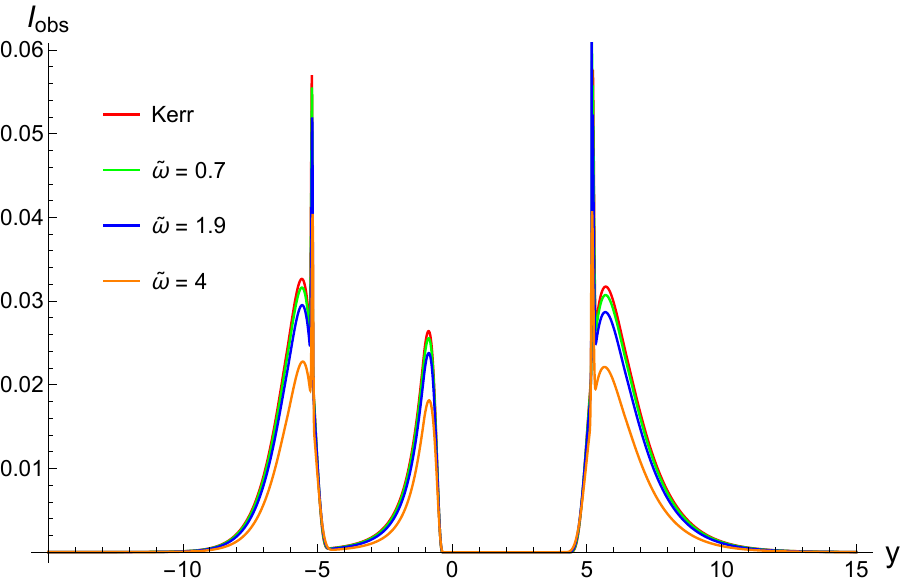}
        \vspace{0.5em}
        \tiny{(d) $a = 0.1$}
    \end{minipage}
    \hfill
    \begin{minipage}{0.32  \textwidth}
        \centering
        \includegraphics[width=\linewidth]{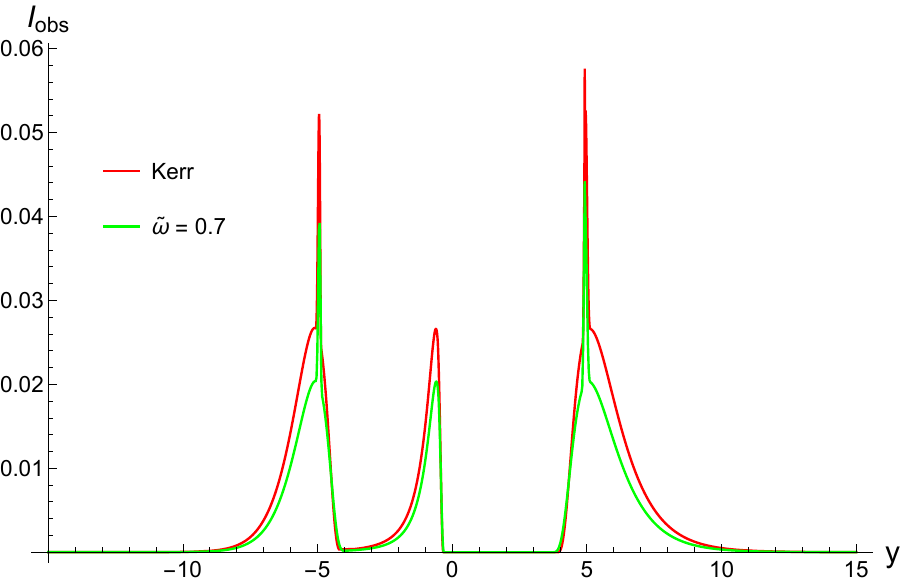}
        \vspace{0.5em}
        \tiny{(e) $a = 0.8$}
    \end{minipage}
    \hfill
    \begin{minipage}{0.32 \textwidth}
        \centering
        \includegraphics[width=\linewidth]{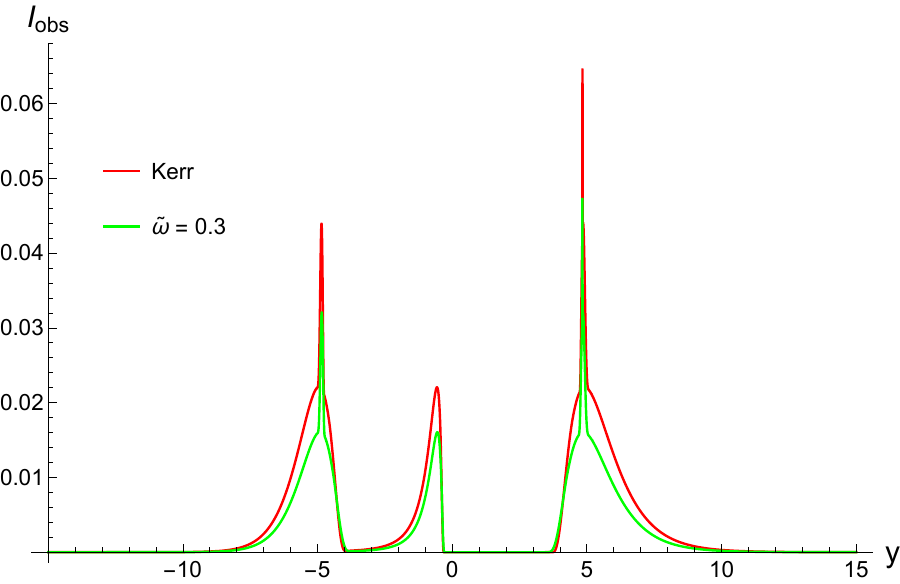}
        \vspace{0.5em}
        \tiny{(f) $a = 0.9$}
    \end{minipage}
    \caption{The observed intensity distribution along x-axis (upper panel) and y-axis (bottom panel) for selected parameters.}
\label{xyintensityplot}
\end{figure}

Figure~\ref{intensityplot} shows that
the images of QIRK black holes illuminated by a thin accretion disk exhibit several notable features.  
The central dark region,  
known as the inner shadow,  
corresponds to the direct image of the event horizon.  
Outside this region,  
the brightest ring structure is the photon ring,  
formed by unstable spherical photon orbits.  
It can be observed that as the spin parameter \(a\) increases,  
the photon ring becomes more distorted and deviates further from a circle.  
The up-down asymmetry of the image is largely determined by the observer's inclination angle,  
while the left-right asymmetry arises mainly from the rotation of the black hole and the orbital motion of the accreting matter.
Our results show that the quantum correction parameter \(\widetilde{\omega}\)  
produces only modest changes in the size and shape of the inner shadow and photon ring. 
However, as \(\widetilde{\omega}\) increases,  
the most pronounced effect is a systematic decrease in overall image brightness (Fig.~\ref{xyintensityplot}),  
present for all spins \(a\).
The emissivity profile $I_{\mathrm{em}}(r)$ in (\ref{emmisionprofile}) depends on the outer horizon $r_h$, 
and $r_h$ decreases as \(\widetilde{\omega}\) grows. 
Because photon trajectories are only weakly affected by \(\widetilde{\omega}\), a given ray intersects the disk at nearly the same radius, but a smaller $r_h$ reduces $I_{\mathrm{em}}(r)$ at fixed $r$, yielding lower local emissivity for larger \(\widetilde{\omega}\).  
The redshift factor $g_n$ varies only mildly--
indeed, it increases slightly as \(\widetilde{\omega}\) grows--
but this is insufficient to compensate for the emissivity suppression. 
Consequently, the observed specific intensity $g_n^3 I_{\mathrm{em}}$ decreases systematically with increasing \(\widetilde{\omega}\).

\begin{figure}[htbp]
    \begin{minipage}{0.22 \textwidth}
        \centering
        \includegraphics[width=\linewidth]{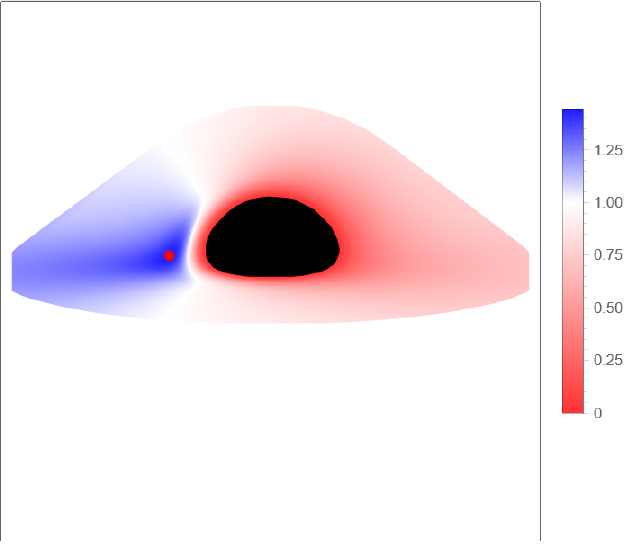}
        \vspace{0.5em}
        \tiny{(a) Kerr $a = 0.1$}
    \end{minipage}
    \hfill
    \begin{minipage}{0.22 \textwidth}
        \centering
        \includegraphics[width=\linewidth]{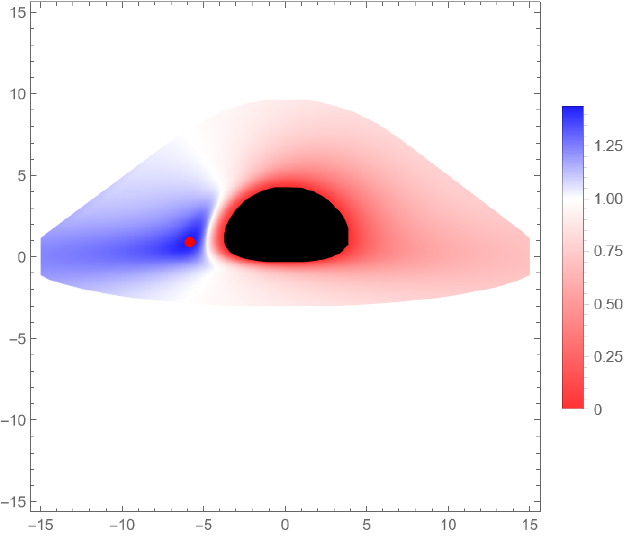}
        \vspace{0.5em}
        \tiny{(b) $a = 0.1$ $p = 4$ $\widetilde{\omega} = 0.7$}
    \end{minipage}
    \hfill
    \begin{minipage}{0.22 \textwidth}
        \centering
        \includegraphics[width=\linewidth]{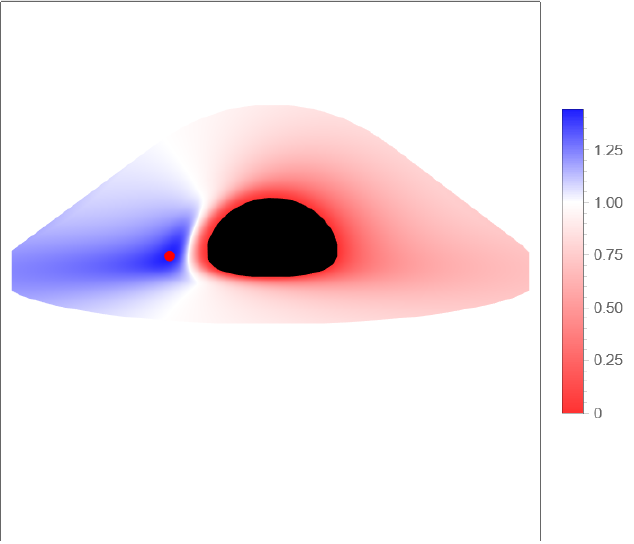}
        \vspace{0.5em}
        \tiny{(c) $a = 0.1$ $p = 4$ $\widetilde{\omega} = 1.9$}
    \end{minipage}
    \hfill
    \begin{minipage}{0.22 \textwidth}
        \centering
        \includegraphics[width=\linewidth]{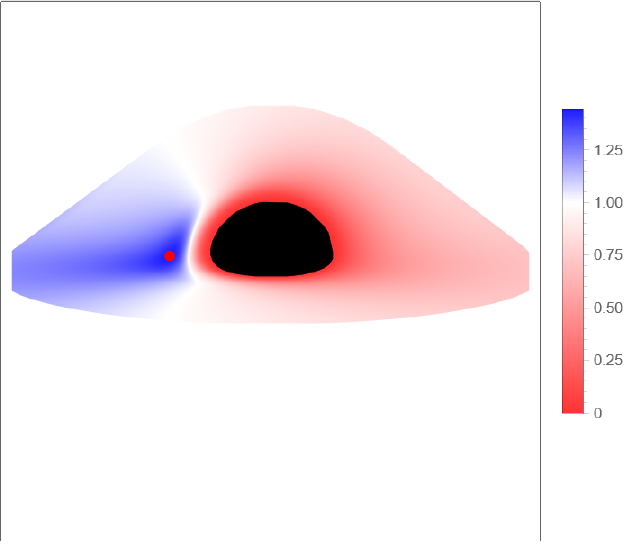}
        \vspace{0.5em}
        \tiny{(d) $a = 0.1$ $p = 4$ $\widetilde{\omega} = 4$}
    \end{minipage}
    \hfill
    \begin{minipage}{0.22  \textwidth}
        \centering
        \includegraphics[width=\linewidth]{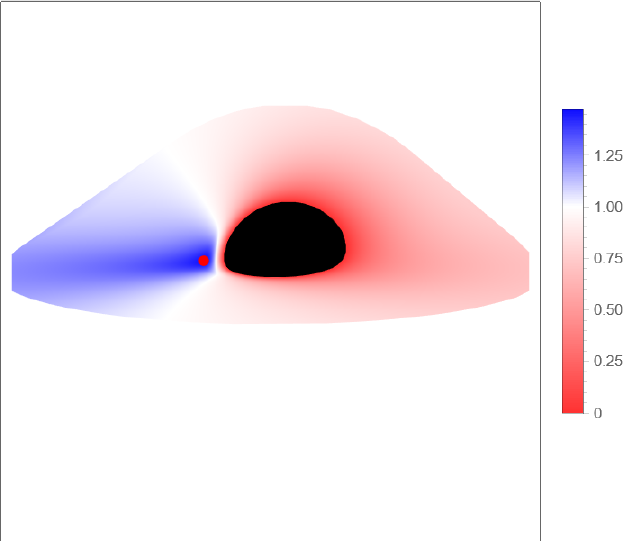}
        \vspace{0.5em}
        \tiny{(e) Kerr $a = 0.8$}
    \end{minipage}
    \hfill
    \begin{minipage}{0.22 \textwidth}
        \centering
        \includegraphics[width=\linewidth]{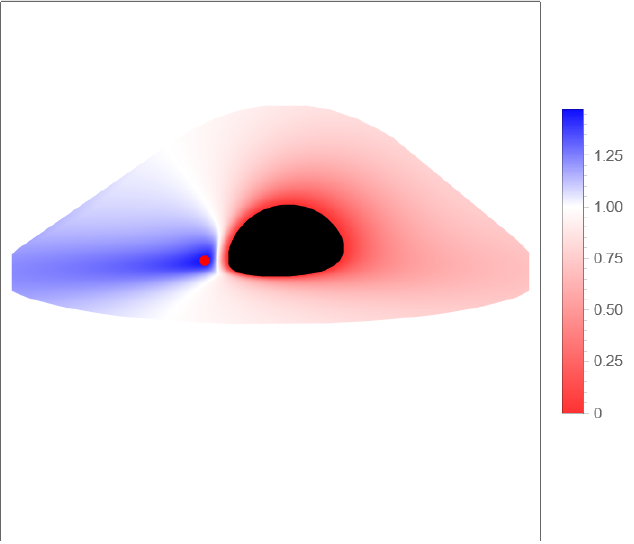}
        \vspace{0.5em}
        \tiny{(f) $a = 0.8$ $p = 4$ $\widetilde{\omega} = 0.7$}
    \end{minipage}
    \hfill
    \begin{minipage}{0.22 \textwidth}
        \centering
        \includegraphics[width=\linewidth]{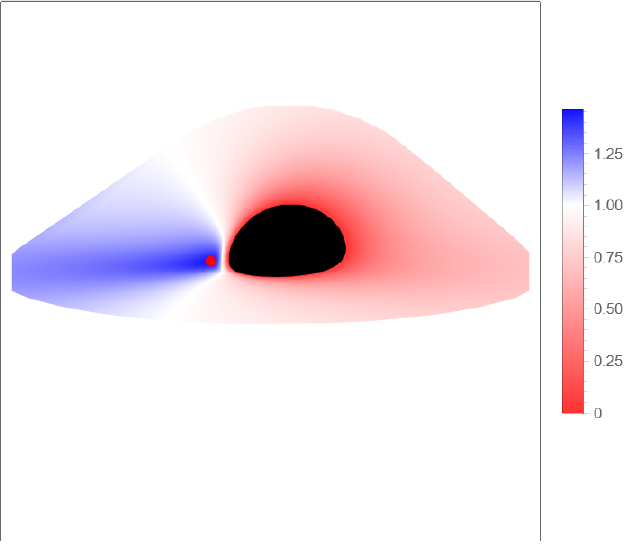}
        \vspace{0.5em}
        \tiny{(g) Kerr $a = 0.9$}
    \end{minipage}
    \hfill
    \begin{minipage}{0.22 \textwidth}
        \centering
        \includegraphics[width=\linewidth]{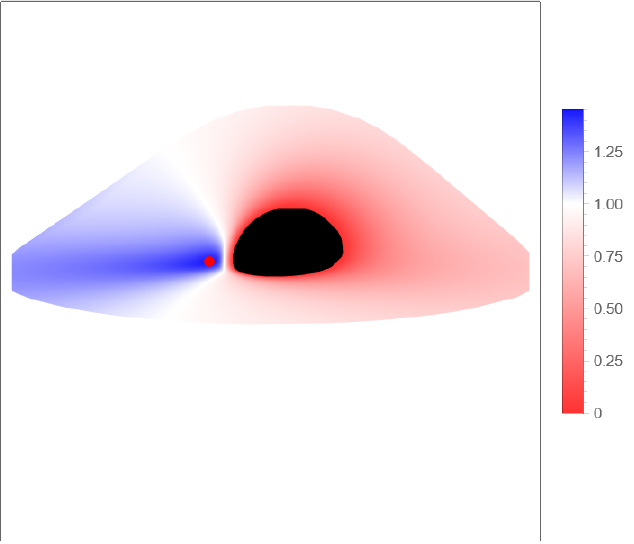}
        \vspace{0.5em}
        \tiny{(h) $a = 0.9$ $p = 4$ $\widetilde{\omega} = 0.3$}
    \end{minipage}
    \caption{The redshift factor distributions of the direct images of the accretion disk. The red and blue indicate redshift and
    blueshift, respectively. And the red dot in each plot indicates the maximal blueshift point.   
    }
\label{directredshiftplot}
\end{figure}

\begin{figure}[htbp]
    \begin{minipage}{0.22 \textwidth}
        \centering
        \includegraphics[width=\linewidth]{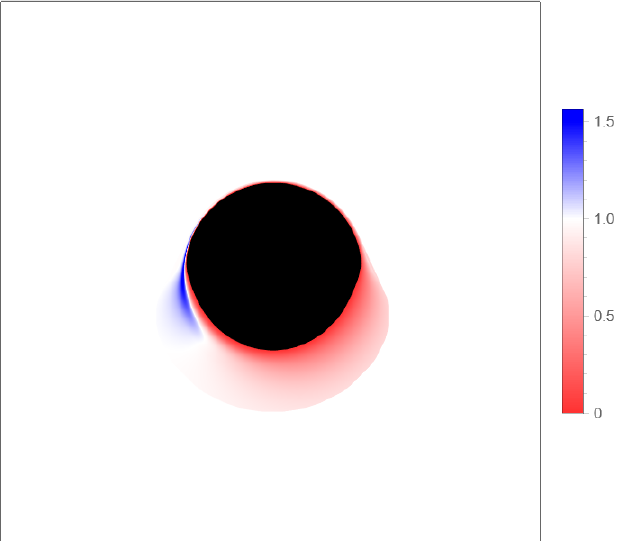}
        \vspace{0.5em}
        \tiny{(a) Kerr $a = 0.1$}
    \end{minipage}
    \hfill
    \begin{minipage}{0.22 \textwidth}
        \centering
        \includegraphics[width=\linewidth]{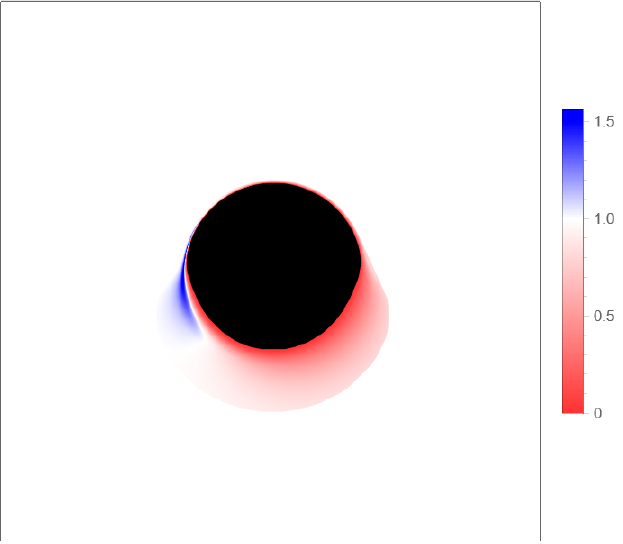}
        \vspace{0.5em}
        \tiny{(b) $a = 0.1$ $p = 4$ $\widetilde{\omega} = 0.7$}
    \end{minipage}
    \hfill
    \begin{minipage}{0.22 \textwidth}
        \centering
        \includegraphics[width=\linewidth]{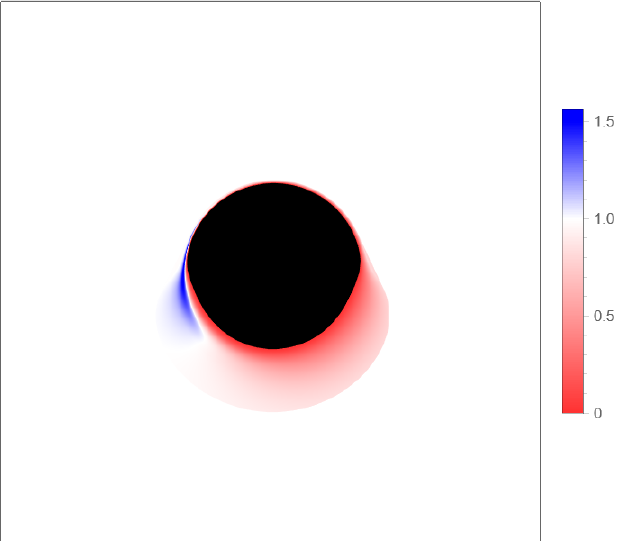}
        \vspace{0.5em}
        \tiny{(c) $a = 0.1$ $p = 4$ $\widetilde{\omega} = 1.9$}
    \end{minipage}
    \hfill
    \begin{minipage}{0.22 \textwidth}
        \centering
        \includegraphics[width=\linewidth]{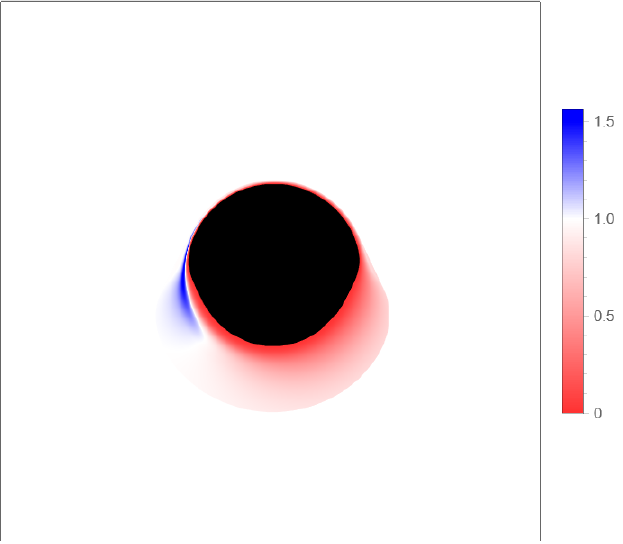}
        \vspace{0.5em}
        \tiny{(d) $a = 0.1$ $p = 4$ $\widetilde{\omega} = 4$}
    \end{minipage}
    \hfill
    \begin{minipage}{0.22  \textwidth}
        \centering
        \includegraphics[width=\linewidth]{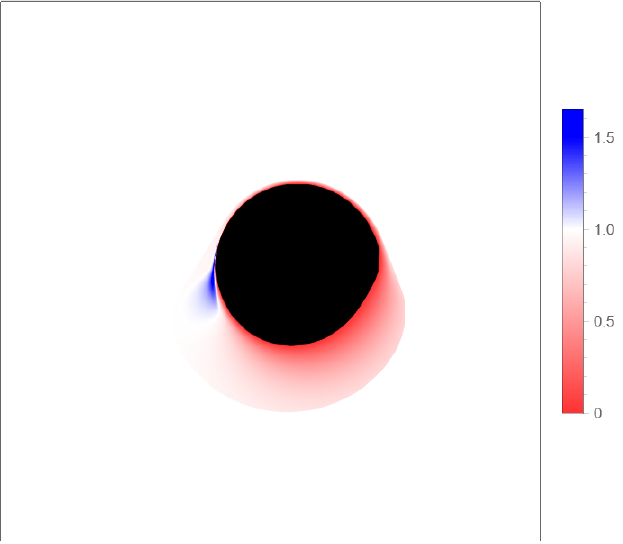}
        \vspace{0.5em}
        \tiny{(e) Kerr $a = 0.8$}
    \end{minipage}
    \hfill
    \begin{minipage}{0.22 \textwidth}
        \centering
        \includegraphics[width=\linewidth]{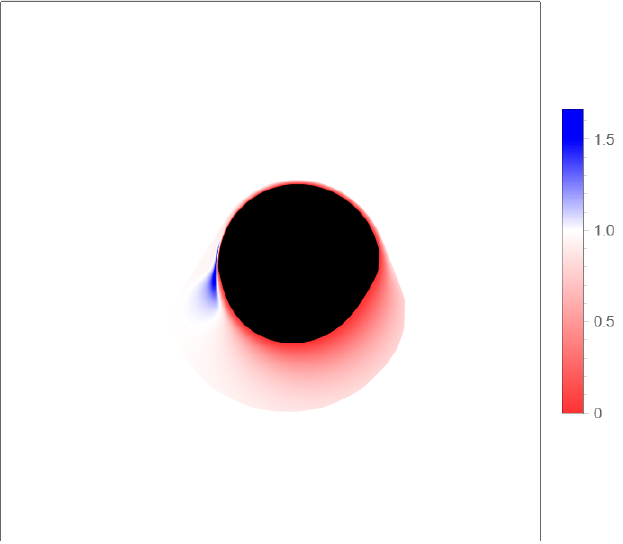}
        \vspace{0.5em}
        \tiny{(f) $a = 0.8$ $p = 4$ $\widetilde{\omega} = 0.7$}
    \end{minipage}
    \hfill
    \begin{minipage}{0.22 \textwidth}
        \centering
        \includegraphics[width=\linewidth]{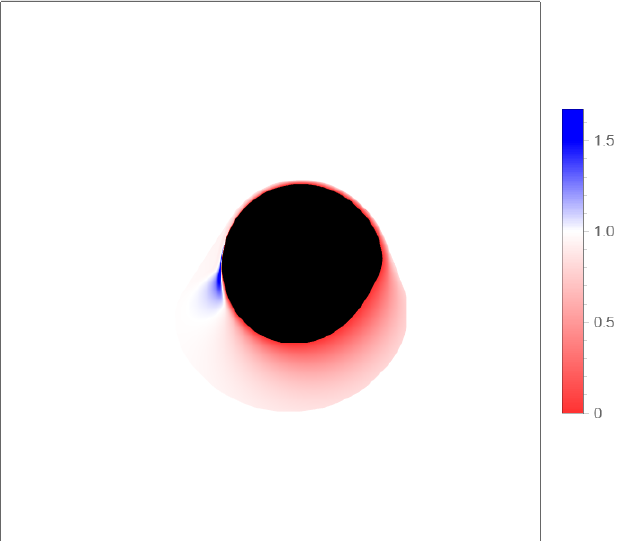}
        \vspace{0.5em}
        \tiny{(g) Kerr $a = 0.9$}
    \end{minipage}
    \hfill
    \begin{minipage}{0.22 \textwidth}
        \centering
        \includegraphics[width=\linewidth]{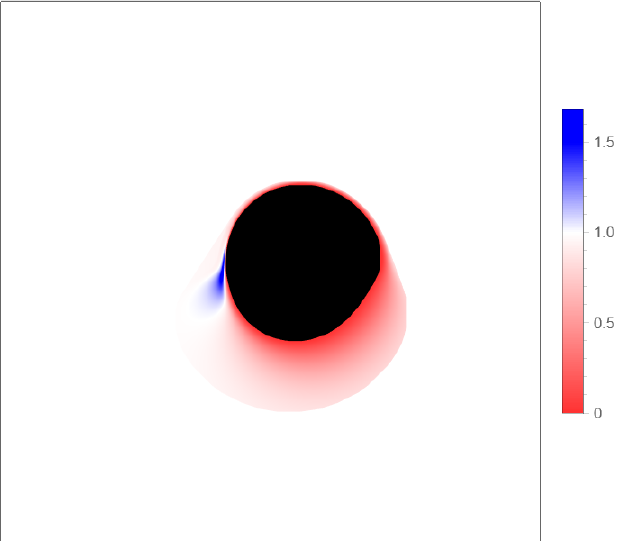}
        \vspace{0.5em}
        \tiny{(h) $a = 0.9$ $p = 4$ $\widetilde{\omega} = 0.3$}
    \end{minipage}
    \caption{ The redshift factors distribution of the lensed images of the accretion disk. The red and blue indicate redshift and
    blueshift, respectively. And the border of the black region in each plot gives the lensed images of event horizon.}
\label{lensedredshiftplot}
\end{figure}

Fig.~\ref{directredshiftplot} and~\ref{lensedredshiftplot} display the redshift distributions of the direct image (\(n=1\)) and the lensed image (\(n=2\)) of the accretion disk.  
The plots show several features.  
The redshift includes contributions from both gravitational redshift and Doppler shift.  
On the left side of each plot, a noticeable blueshift appears due to the prograde motion of the accretion flow.  
If the flow were retrograde, the pattern would change significantly.  
As a result, the overall redshift map is dominated by gravitational redshift as a background, with Doppler effects providing the main variation.  
Under the current observational setup and parameter choices,  
quantum corrections cause only minor changes to this macroscopic appearance, including the magnitude of the redshift factor \(g_n\).  
This means that distinguishing QIRK black holes with different \(\widetilde{\omega}\),  
or separating QIRK from Kerr black hole based solely on the large scale morphology of the redshift map,  
is likely to be difficult.  
A more reliable way to distinguish them will require a precise quantitative analysis of geometric observables,  
which we discuss in the next section.

\section{Constraints from EHT Observations}\label{Constraining with EHT Observations}

The previous analysis of QIRK black hole images suggests that quantum corrections induce only subtle modifications to the main features, 
manifesting primarily as minor shifts in intensity or geometry.  
Although these qualitative findings are very useful,  
a more rigorous quantitative analysis is required for precise characterization.  
Certain observables that describe the geometric features of the shadow  
provide this capability.   
In this section,  
we analyze three geometric observables of the black hole shadow:  
the circularity deviation \(\Delta C\),  
the shadow angular diameter \(\theta_d\),  
and the Schwarzschild deviation \(\delta\).  
Assuming that the supermassive black holes M87* and Sgr~A* are described by the QIRK model,  
we use Event Horizon Telescope (EHT) observations to constrain the QIRK parameters.  
Through this procedure,  
we aim to investigate the quantum corrections introduced by QIRK in greater detail,  
and align the QIRK model more closely with astronomical data.

The black hole shadow boundary is characterized as a one-dimensional closed curve described by radial and angular coordinates $(R(\varphi), \varphi)$ in a polar coordinate system centered at $(x_C, y_C)$. 
The shadow's average radius $\bar{R}$ is defined as follows
\cite{johannsen_testing_2010}:
\begin{align}
    \bar{R} = \frac{1}{2\pi} \int_{0}^{2\pi} R(\varphi) d\varphi, \label{barR}
\end{align}
with
\begin{align}
    R(\varphi) = \sqrt{(x - x_C)^2 + (y - y_C)^2}, \quad \tan(\varphi) = \frac{y - y_C}{x - x_C},
\end{align}
where $(x_C, y_C)$ represent the shadow's displacement from the black hole center at $(0, 0)$. 
Given the intrinsic axisymmetry, 
the vertical displacement is zero ($y_C = 0$), 
and the horizontal displacement $x_C$ is
\begin{align}
    x_C = \frac{x_r + x_l}{2},
\end{align}
where $x_r$ and $x_l$ denote the abscissae where the shadow intersects the $x$-axis.

The circularity deviation $\Delta C$, 
measuring the deviation from a perfect circle, 
is defined as the root-mean-square distance from the average radius
\cite{johannsen_testing_2010, johannsen_photon_2013, walia_testing_2022}:
\begin{align}
    \Delta C = 2\sqrt{\frac{1}{2\pi} \int_{0}^{2\pi} (R(\varphi) - \bar{R})^2 d\varphi},
\end{align}
where $\Delta C = 0$ for a perfectly circular shadow. 
From EHT observations, 
the circularity deviation for M87* has been constrained to $\Delta C \leq 0.10$
\cite{EventHorizonTelescope:2019dse, EventHorizonTelescope:2019ggy, EventHorizonTelescope:2019pgp}.

The shadow's angular diameter $\theta_d$, 
defined in terms of the shadow cone's opening angle, 
is given by~\cite{afrin_parameter_2021}:
\begin{align}
    \theta_d = \frac{2}{D_{LS}} \sqrt{\frac{A}{\pi}},
\end{align}
where $A$ is the shadow area, expressed as~\cite{abdujabbarov_coordinate-independent_2015}:
\begin{align}
    A = 2 \int y(r_p) dx(r_p) = 2 \int_{r^-_p}^{r^+_p} \left(y(r_p) \frac{dx(r_p)}{r_p}\right) dr_p.
\end{align}

Additionally, 
the observable $\delta$, 
quantifying the deviation between the observed angular diameter and that of a Schwarzschild black hole, 
is defined as
\cite{EventHorizonTelescope:2022xqj, EventHorizonTelescope:2022urf}:
\begin{align}
    \delta = \frac{\theta_d}{\theta_{d,Sch}} - 1,
\end{align}
where $\theta_{d,Sch}$ is the angular diameter of a Schwarzschild black hole's shadow.

In the following, 
we aim to constrain the model parameters based on astronomical data.
By analyzing these observables, 
we will exclude parameter ranges incompatible with the data, 
using density maps based on Fig.~\ref{range_changep}. 
While uncertainties in the EHT observations of M87* and Sgr A* introduce some degree of imprecision, 
we can still estimate the approximate effect of the parameters introduced by the QIRK black holes in the observations.

The viewing inclination generally differs from source to source.
For instance, EHT estimates the inclination angle of M87* to be approximately $163^\circ$, 
based on the relativistic jet orientation
\cite{CraigWalker:2018vam}. 
Since the shadow is symmetric about the $x$-axis, $\theta_0 = 163^\circ$ is equivalent to $\theta_0 = 17^\circ$ in our analysis. 
For Sgr A*, while the inclination angle is not definitively known, 
values below $52^\circ$ are generally favored. 
Inclinations of $5^\circ$~\cite{Kuang:2022ojj}, $50^\circ$~\cite{walia_testing_2022}, and $90^\circ$~\cite{Ghosh:2022kit} 
have been employed in previous analyses. 
The circularity deviation \( \Delta C \) increases with the inclination angle \( \theta_0 \), 
so we can obtain an upper bound on the parameters by choosing an inclination of \( 90^\circ \). 
However, the shadow area decreases as the inclination angle increases, 
and the angular diameter \( \theta_d \) and Schwarzschild deviation \( \delta \) are proportional to the area. 
Therefore, smaller inclination angles can provide tighter constraints when considering \( \theta_d \) and \( \delta \). 
After taking these factors into account, 
we select \( \theta_0 = 17^\circ \) and \( 90^\circ \) for M87*, and \( 50^\circ \) for Sgr A*.

\subsection{Constraints from M87*}

The first image of the supermassive black hole M87* revealed an asymmetric bright ring caused by strong gravitational lensing and relativistic beaming, 
with a central dark region identified as the black hole shadow
\cite{EventHorizonTelescope:2019dse, EventHorizonTelescope:2019ggy, EventHorizonTelescope:2019pgp}. 
Given the distance of M87* from Earth, $D_{ \text{M87*} } = 16.8$ Mpc, 
and its estimated mass $M_{\text{M87*}} = (6.5 \pm 0.7) \times 10^9 M_\odot$, 
constraints can be placed on the emission region. 
The circularity deviation is constrained to $\Delta C \leq 0.10$, 
with an angular diameter of $\theta_{d,\text{M87*}} = (42 \pm 3) \mu \text{as}$ and a Schwarzschild deviation of $\delta_{\text{M87*}} = -0.01 \pm 0.17$ within a $1\sigma$ confidence interval.

\begin{figure}[!ht]
    \centering
    \includegraphics[width=5.9cm]{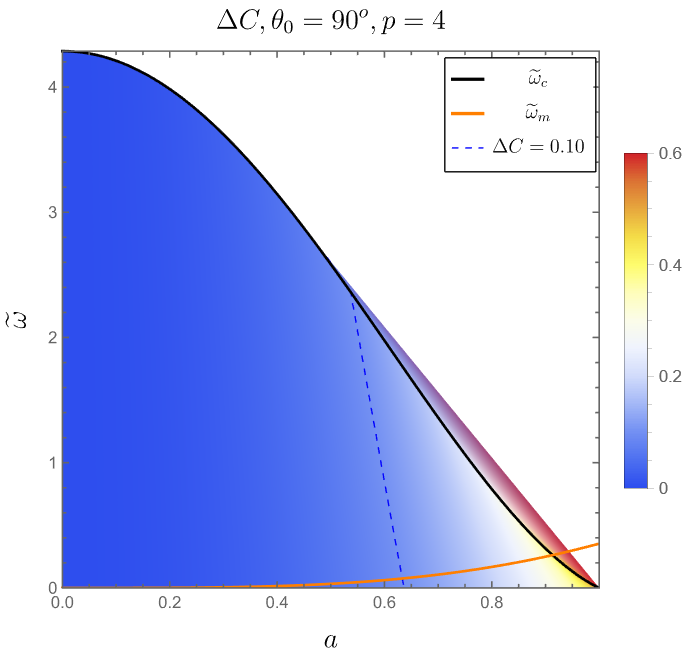}
    \includegraphics[width=5.9cm]{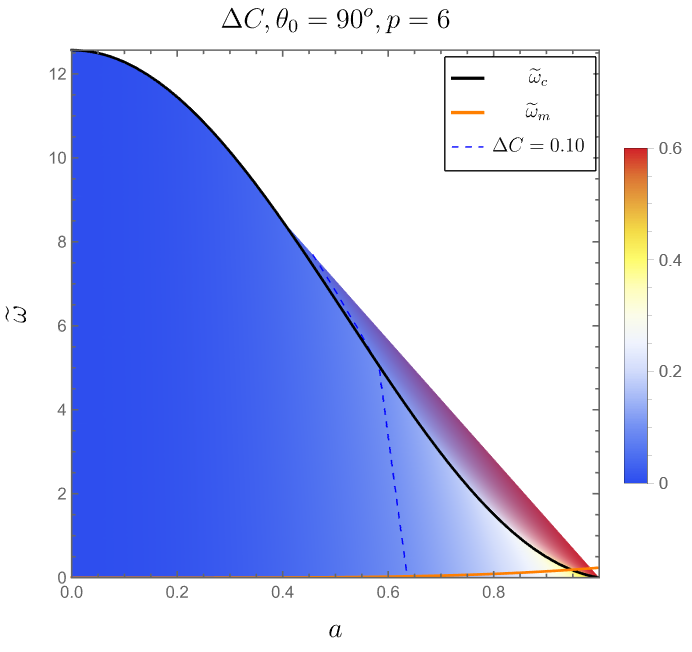}
    \includegraphics[width=5.9cm]{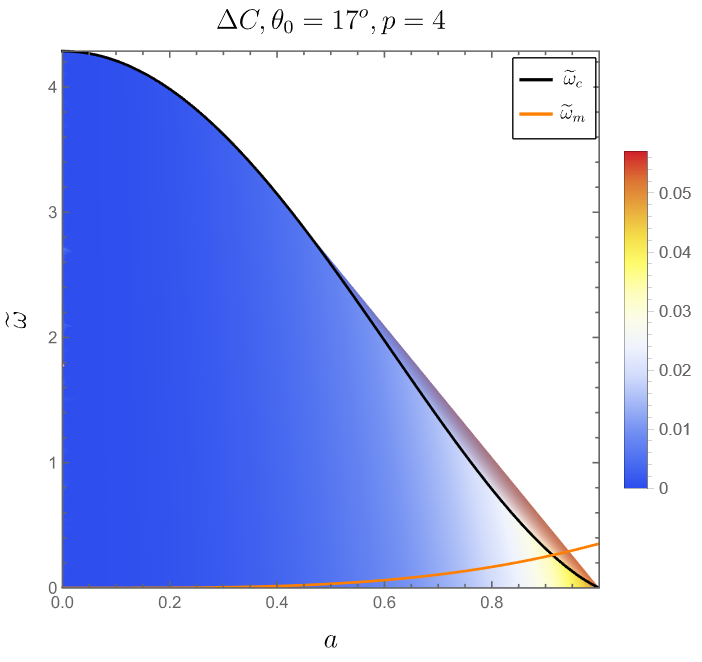}
    \caption{
        The shadow circularity deviation observable $\Delta C$ for different inclination angles $\theta_0$ and parameter $p$, 
        as a function of $(a, \widetilde{\omega})$. 
        The dashed blue curve indicates $\Delta C = 0.10$, 
        where the region to the right of this curve is excluded based on the observed circularity deviation of the M87* black hole as reported by the EHT, 
        $\Delta C \leq 0.10$.}
    \label{M87_DeltaC}
\end{figure}

\begin{figure}[!ht]
    \centering
    \includegraphics[width=5.9cm]{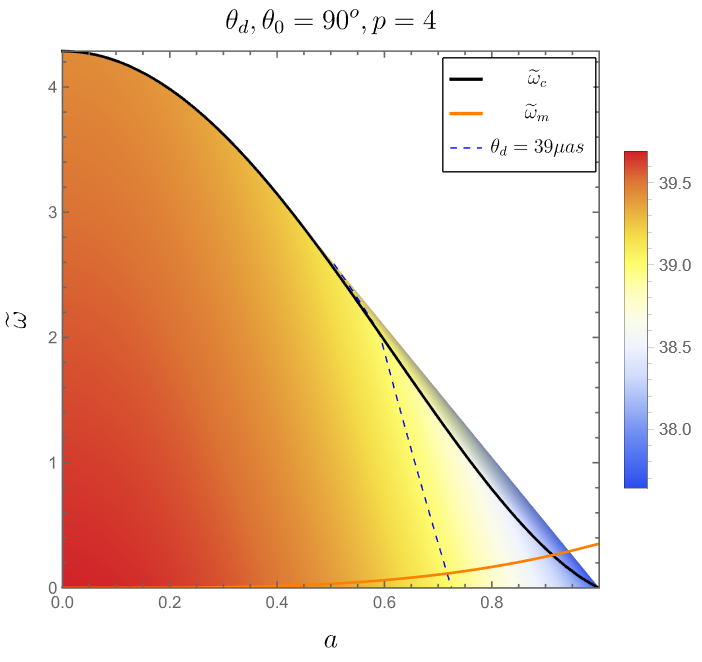}
    \includegraphics[width=5.9cm]{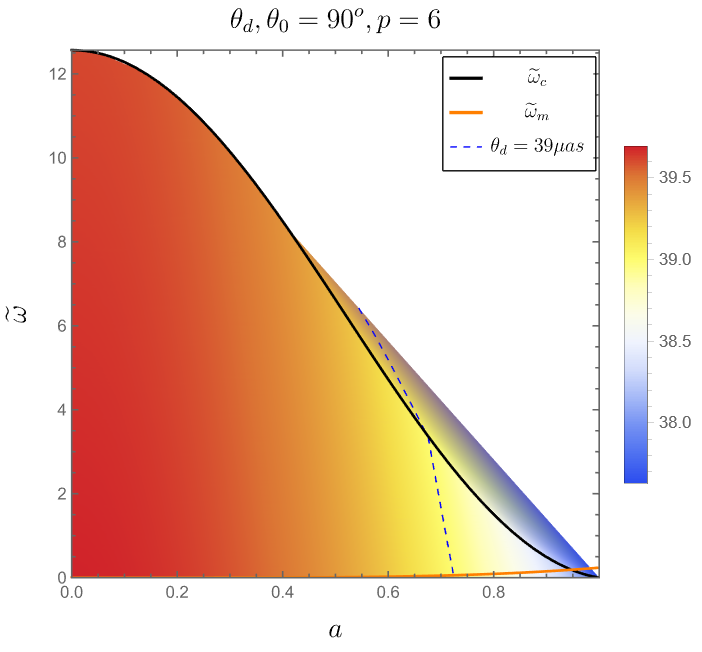}
    \includegraphics[width=5.9cm]{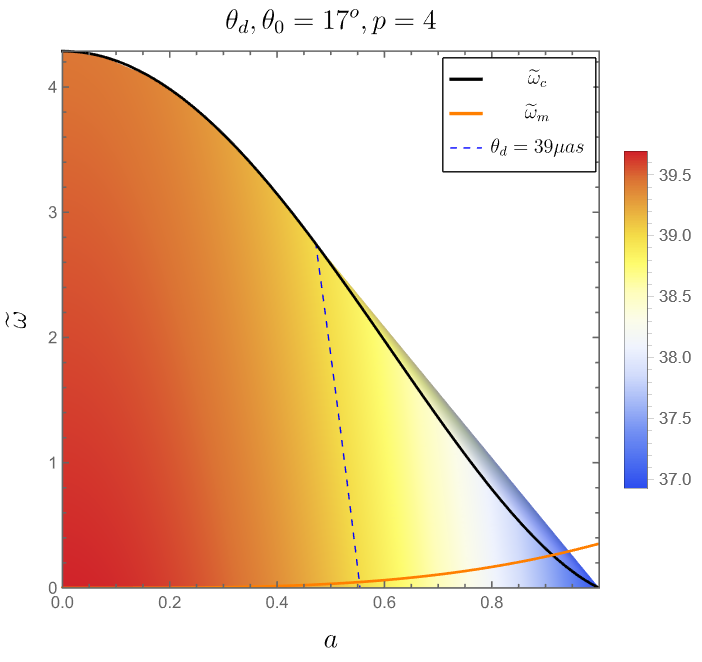}
    \caption{
        The shadow angular diameter \( \theta_d \) for QIRK black holes, 
        with different inclination angles \( \theta_0 \) and parameter \( p \), 
        as a function \( (a, \widetilde{\omega}) \). 
        For M87*, EHT observations report an angular diameter of \( \theta_d = 42 \pm 3 \, \mu \text{as} \). 
        The dashed blue curve corresponds to \( \theta_d = 39 \, \mu \text{as} \), 
        and the region to the right of this curve exceeds the $1\sigma$ range of the observational result, 
        thus being excluded.
        }
    \label{M87_theta}
\end{figure}

\begin{figure}[!ht]
    \centering
    \includegraphics[width=5.9cm]{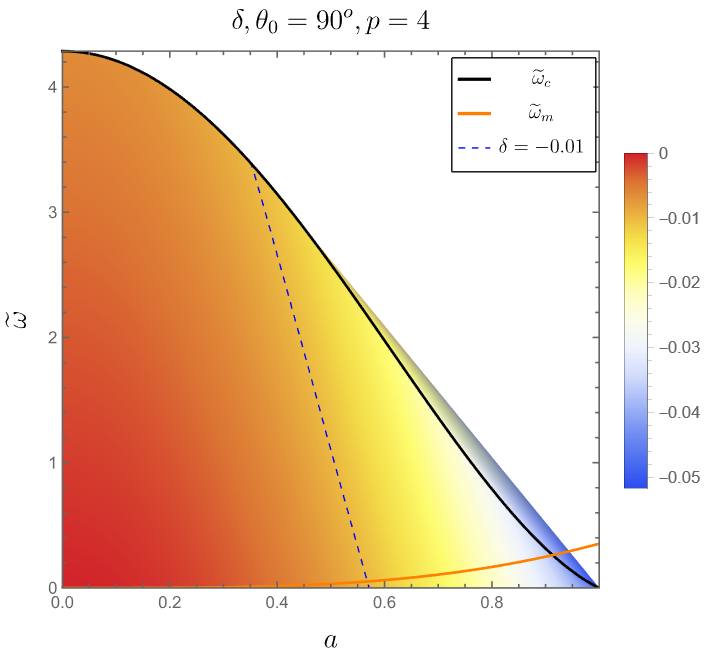}
    \includegraphics[width=5.9cm]{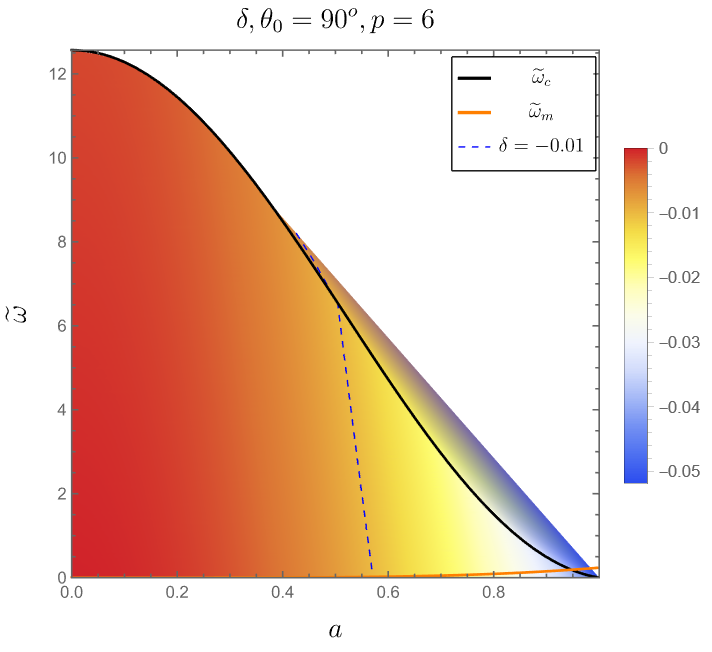}
    \includegraphics[width=5.9cm]{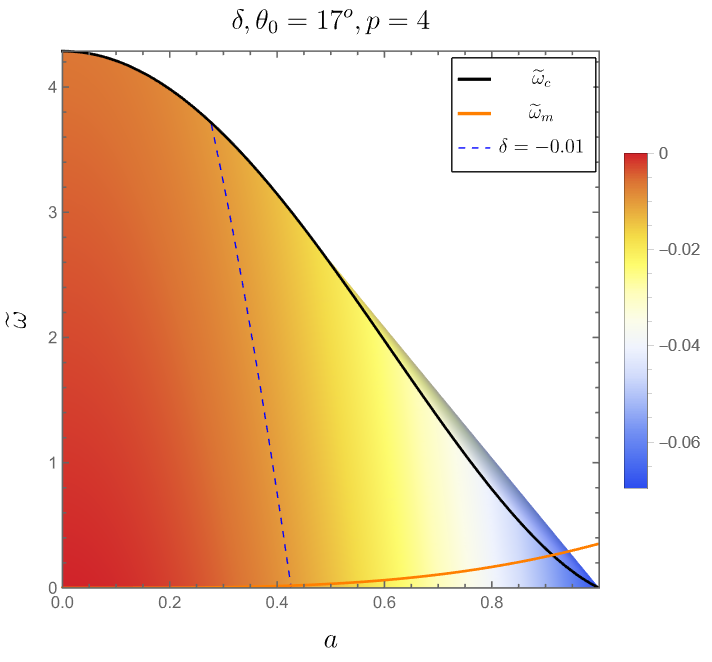}
    \caption{
        Schwarzschild shadow angular diameter deviation $\delta$ as a function of $(a, \widetilde{\omega})$. 
        The dashed blue curve represents $\delta = -0.01$.}
    \label{M87_sdelta}
\end{figure}

Figs.~\ref{M87_DeltaC}, \ref{M87_theta}, and \ref{M87_sdelta} display the density plots of $\Delta C$, $\theta_d$, and $\delta$ under varying inclinations and parameters. 
As shown in Fig.~\ref{M87_DeltaC}, $\Delta C$ remains below 0.10 for $\theta_0 = 17^\circ$, providing no substantial constraints. 
However, for $\theta_0 = 90^\circ$, 
some parameters with $\Delta C > 0.10$ are excluded.
From Fig.~\ref{M87_theta}, 
we can observe that the area of the shadow \( A \) increases as the inclination angle \( \theta_0 \) decreases.
Therefore, $\theta_0 = 17^\circ$ provides stronger constraints if we exclude $\theta_d < 39 \mu \text{as}$. For $\theta_d$, 
the exclusion criteria are $\theta_d > 45 \mu \text{as}$ or $\theta_d < 39 \mu \text{as}$ within the $1\sigma$ interval. 
Given the modest deviation of the QIRK black hole from the Kerr black hole, 
we do not extend the analysis to the $2\sigma$ interval. 
Finally, Fig.~\ref{M87_sdelta} shows that all values of $\delta$ lie within the $1\sigma$ interval, 
meaning no significant exclusion can be made based on $\delta$ alone.

\subsection{Constraints from Sgr A*}

In the published results for Sgr A*, 
the EHT provides measurements for both the angular diameter of the emission ring, 
$\theta_{d,\text{Sgr~A*}} = (51.8 \pm 2.3) \mu\text{as}$, 
and the angular diameter of the black hole shadow, 
$\theta_{sh,\text{Sgr~A*}} = (48.7 \pm 7) \mu\text{as}$. 
Given the broad range of the shadow angular diameter, 
we will use an average value of $\theta_d \in (46.9, 50) \mu\text{as}$, 
as determined by the three independent imaging algorithms employed by the EHT~\cite{walia_testing_2022}. 
For uniformity, we will denote all shadow angular diameters simply as $\theta_d$.
The Schwarzschild shadow deviation is given by $\delta_{\text{Sgr~A*}} = -0.08^{+0.09}_{-0.09}$ (VLTI) and $\delta_{\text{Sgr~A*}} = -0.04^{+0.09}_{-0.10}$ (Keck) at the $1\sigma$ confidence level. 
In the following analysis, 
we adopt the mass of Sgr A* as $M_{\text{Sgr~A*}} = 4.0^{+1.1}_{-0.6} \times 10^6 M_\odot$ and its distance from Earth as $D_{\text{Sgr~A*}} = 8.15 \pm 0.15$ kpc
\cite{EventHorizonTelescope:2022wkp,EventHorizonTelescope:2022xqj,EventHorizonTelescope:2022urf, EventHorizonTelescope:2022wok,EventHorizonTelescope:2022apq,EventHorizonTelescope:2022exc}.

\begin{figure}[!ht]
    \centering
    \includegraphics[width=6.5cm]{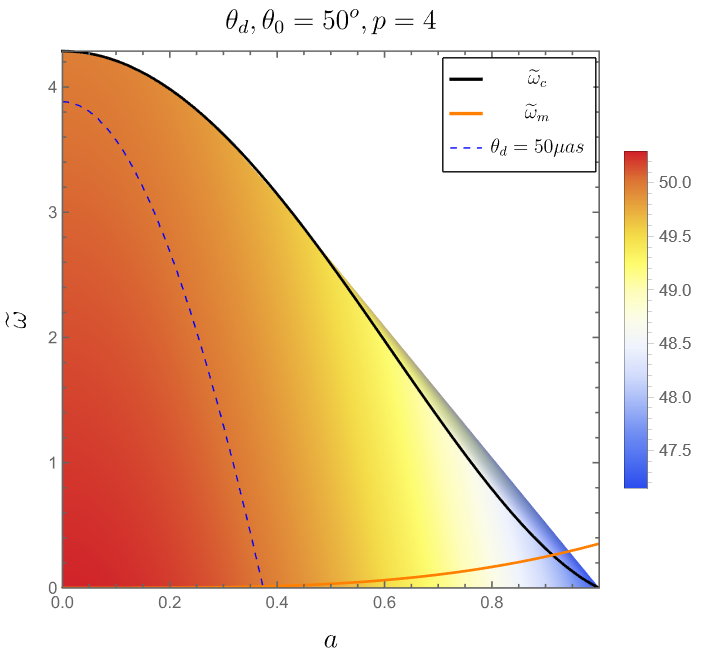}
    \hfil
    \includegraphics[width=6.5cm]{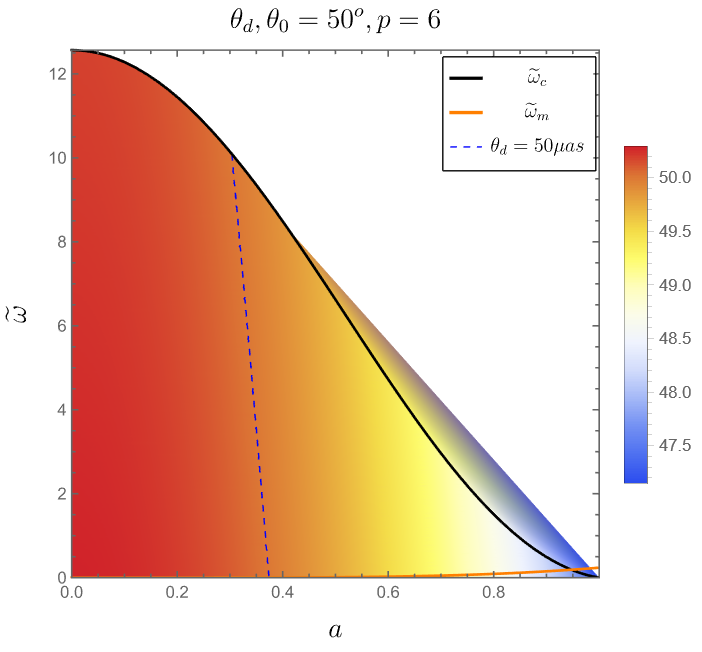}
    \caption{
        Shadow angular diameter $\theta_d$ for QIRK black holes 
        with different values of parameter $p$, plotted as a function of $(a, \widetilde{\omega})$. 
        The dashed blue curve represents $\theta_d = 50 \mu \text{as}$, corresponding to the Sgr A* black hole shadow bounds. 
        QIRK black holes in the region to the right of the blue curve produce shadows consistent with the observed size of the Sgr A* shadow.}
    \label{SgrA_theta}
\end{figure}

\begin{figure}[!ht]
    \centering
    \includegraphics[width=6.5cm]{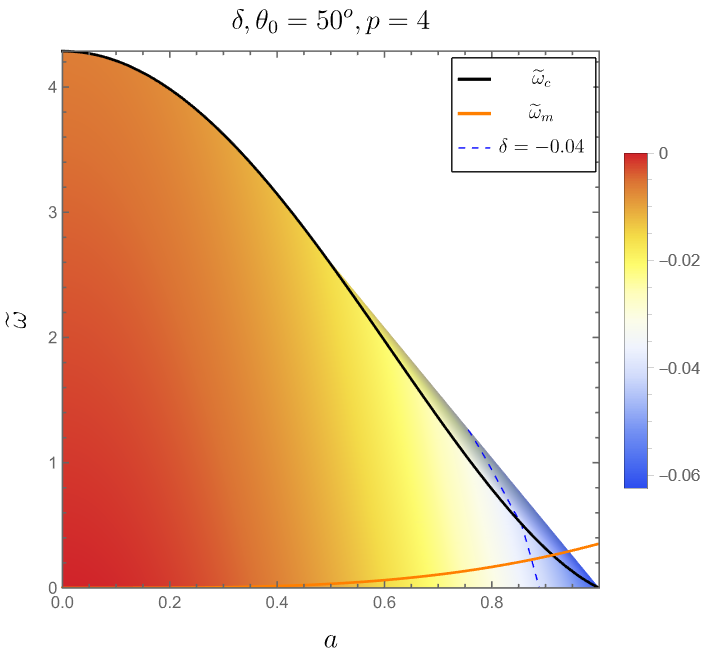}
    \hfil
    \includegraphics[width=6.5cm]{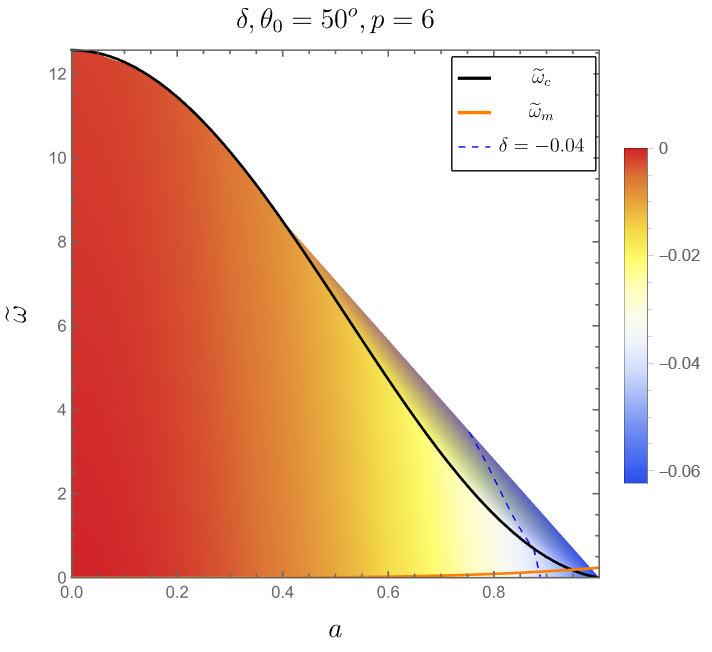}
    \caption{
        Schwarzschild shadow angular diameter deviation $\delta$ as a function of $(a, \widetilde{\omega})$. 
        The dashed blue curve corresponds to $\delta = -0.04$ (Keck).}
    \label{SgrA_sdelta}
\end{figure}

For Sgr A*, 
only the angular diameter $\theta_d$ and deviation $\delta$ are considered, 
as the EHT did not provide a bound on the circularity deviation $\Delta C$. 
As with M87*, 
Fig.~\ref{SgrA_theta} allows us to exclude parameters where $\theta_d > 50 \mu\text{as}$. 
Additionally, Fig.~\ref{SgrA_sdelta} plots $\delta = -0.04$ (Keck), 
but this does not yield meaningful constraints.

\begin{figure}[!ht]
    \centering
    \includegraphics[width=6.3cm]{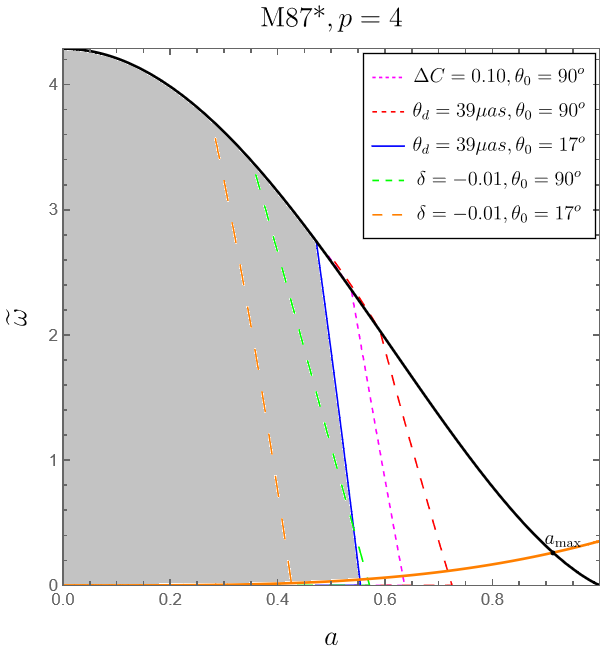}
    \includegraphics[width=6.3cm]{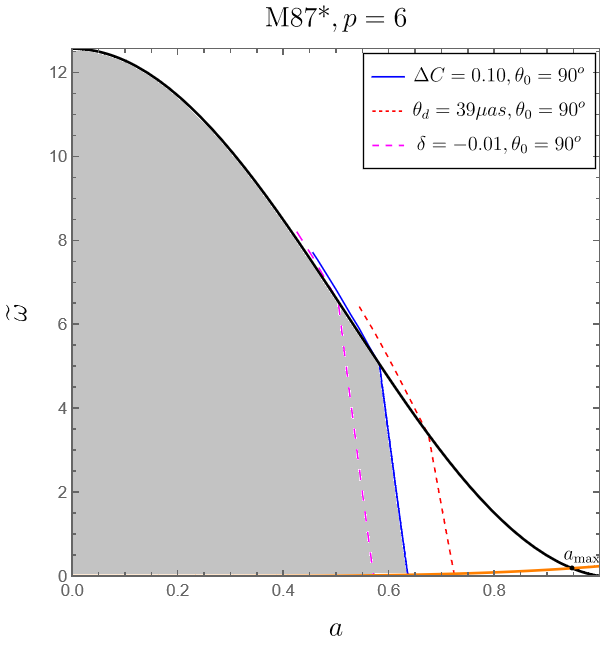}
    \includegraphics[width=6.3cm]{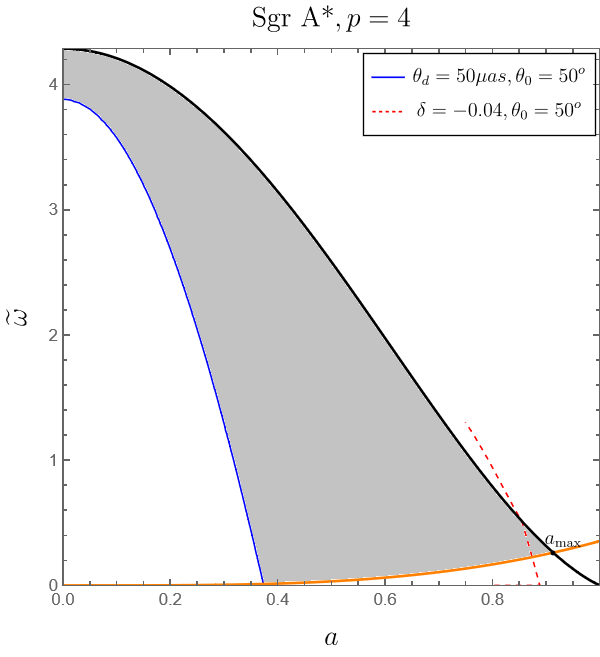}
    \includegraphics[width=6.3cm]{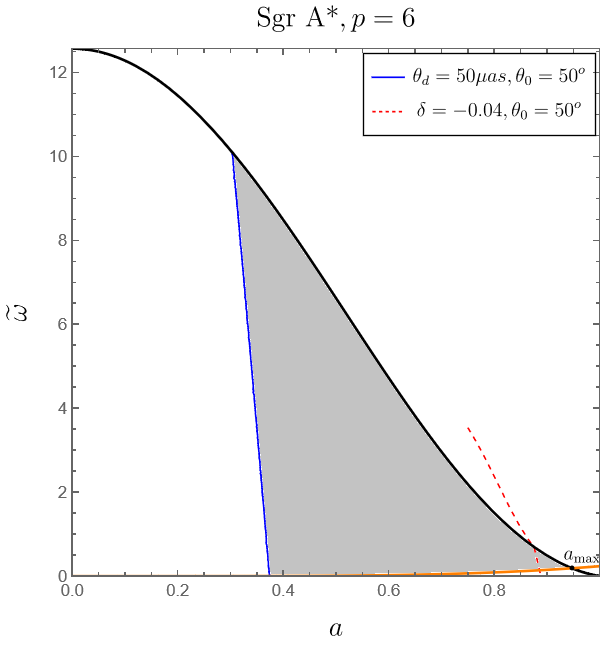}
    \caption{
    The observational bounds for M87* and Sgr A* with different \( p \). 
    The black and orange curves correspond to \(\widetilde{\omega}_c\) and \(\widetilde{\omega}_m\), respectively,  
    as in Fig.~\ref{range_changep}.  
    Their intersection defines \(a_{\text{max}}\),  
    the maximum value of \(a\) for which both the existence of a horizon and the absence of CTCs are ensured,  
    as discussed in Sec.~\ref{quantum improved Regular Kerr Black Hole}.  
    In the first panels, 
    parameters to the right are excluded by the blue curve \( \theta_d = 39 \,\mu \text{as}, \, \theta_0 = 17^\circ \), 
    the red dashed curve \( \theta_d = 39 \,\mu \text{as}, \, \theta_0 = 90^\circ \), 
    and the magenta dashed curve \( \Delta C = 0.10, \, \theta_0 = 90^\circ \). 
    In the second panel, 
    parameters to the right are excluded by the red dashed curve \( \theta_d = 39 \,\mu \text{as}, \, \theta_0 = 90^\circ \) 
    and the blue curve \( \Delta C = 0.10, \, \theta_0 = 90^\circ \). 
    In contrast, 
    in the lower two panels, 
    parameters to the left are excluded by the blue curve \( \theta_d = 50 \,\mu \text{as}, \, \theta_0 = 50^\circ \). 
    In summary, 
    the blue curves give the tightest combined constraints; grey regions remain allowed.
}
    \label{matrix_plot}
\end{figure}

Fig.~\ref{matrix_plot} provides a unified representation of the constraints from all observables. 
For M87*, 
the curve \( \theta_d = 39\,\mu\text{as}, \, \theta_0 = 17^\circ \) excludes a significant number of parameters, 
particularly those to the right of it. 
In contrast, for Sgr A*, 
parameters to the left of \( \theta_d = 50\,\mu\text{as}, \, \theta_0 = 50^\circ \) are excluded. 
In Fig.~\ref{matrix_plot}, the blue curves represent the most stringent constraint on the parameter range,
while the gray regions indicate the ultimately allowed parameter range.

After synthesizing the results from M87* and Sgr A*, 
we obtain a band-like region between \( \theta_d = 39\,\mu\text{as} \), \( \theta_0 = 17^\circ \) 
and \( \theta_d = 50\,\mu\text{as} \), \( \theta_0 = 50^\circ \). 
From this, we observe that for the Kerr black hole, 
there are upper and lower bounds on the rotation parameter \( a \). 
However, the introduction of quantum corrections alters these bounds. 
In general, the larger the correction \( \widetilde{\omega} \), 
the smaller both the upper and lower bounds become. 
Once the correction \( \widetilde{\omega} \) exceeds a certain value, 
the lower bound of \( a \) can even reach zero, 
which is significantly different from the Kerr black hole. 
Moreover, when \( a = 0 \), 
\( \widetilde{\omega} \) is constrained to a region near the \( \widetilde{\omega}_c \), 
with most of its values being excluded.
The precise intersection points between each curve and the \( \widetilde{\omega}_c \), 
\( \widetilde{\omega}_m \), or \( \widetilde{\omega} \)-axis are provided in Table~\ref{M87_table} and Table~\ref{SgrA_table}, 
corresponding to M87* and Sgr A*, respectively.

\begin{table}[htbp]
    \centering
    \caption{Intersection points of observational constraints with the curves \( \widetilde{\omega}_c \), \( \widetilde{\omega}_m \), or the \( \widetilde{\omega} \)-axis for M87*.}
    \begin{tabular}{cccccc|cccc}
        \hline
        \hline
                & \multicolumn{5}{c}{$p=4$} &  \multicolumn{3}{c}{$p=6$}  \\
        \hline
                &$\Delta C=0.1$ & $\theta_d=39 \mu \text{as}$ & $\theta_d=39 \mu \text{as}$ & $\delta =-0.01$ & $\delta =-0.01$ & $\Delta C=0.1$  & $\theta_d=39 \mu \text{as}$  & $\delta =-0.01$ \\
                &$\theta_0=90^\circ$ & $\theta_0=90^\circ$ & $\theta_0=17^\circ$ & $\theta_0=90^\circ$ & $\theta_0=17^\circ$ & $\theta_0=90^\circ$  & $\theta_0=90^\circ$  & $\theta_0=90^\circ$ \\
        \cline{2-6} \cline{7-9}
        $\widetilde{\omega}_c$    & $(0.537,2.361)$ & $(0.591,2.034)$ & $(0.472,2.748)$ & $(0.352,3.386)$ & $(0.277,3.717)$ &  $(0.537,2.361)$ & $(0.675,3.374)$ & $(0.504,6.544)$  \\
        \hline
        $\widetilde{\omega}_m$    & $(0.633,0.075)$ & $(0.717,0.115)$ & $(0.553,0.047)$ & $(0.568,0.051)$ & $(0.426,0.018)$  &$(0.633,0.075)$ & $(0.724,0.058)$ & $(0.571,0.020)$  \\
    \end{tabular}
    \label{M87_table}
\end{table}

\begin{table}[htbp]
    \centering
    \caption{Intersection points of observational constraints with the curves \( \widetilde{\omega}_c \), \( \widetilde{\omega}_m \), or the \( \widetilde{\omega} \)-axis for Sgr A*.}
    \begin{tabular}{c@{\hspace{4 em}}c@{\hspace{4 em}}c@{\hspace{2 em}}|@{\hspace{2 em}}c@{\hspace{4 em}}c}
        \hline
        \hline
                & \multicolumn{2}{c}{$p=4,\theta_0=50^\circ$} &  \multicolumn{2}{c}{$p=6,\theta_0=50^\circ$}  \\
        \hline
                 & $\theta_d=39 \mu \text{as}$      & $\delta =-0.01$    & $\theta_d=39 \mu \text{as}$  & $\delta =-0.01$ \\
        \cline{2-3} \cline{4-5}
        $\widetilde{\omega}_c$ / $\widetilde{\omega}$ -axis   & $(0,3.882)$ & $(0.855,0.513)$ & $(0.304,10.093)$ & $(0.875,0.695)$   \\
        \hline
        $\widetilde{\omega}_m$    & $(0.374,0.011)$ & $(0.874,0.226)$ & $(0.374,0.003)$ & $(0.886,0.141)$   \\
    \end{tabular}
    \label{SgrA_table}
\end{table}

\section{Image of Closed Timelike Curves}\label{Image of Closed Timelike Curves}

Closed timelike curves (CTCs) are pathological features in certain solutions of general relativity,  
most notably in the Kerr spacetime.  
The existence of CTCs is generally confined within the Cauchy horizon (\(r < r_{-}\)),  
where the metric component \(g_{\phi\phi}\) can become negative,  
theoretically violating causality.  
Interestingly,  
the parameter space of the QIRK model reveals cases in which the horizon existence condition \(\widetilde{\omega} \le \widetilde{\omega}_c\) is not met,  
and simultaneously the condition to avoid CTCs \(\widetilde{\omega} > \widetilde{\omega}_m\) is also not satisfied.  
This gives rise to a theoretical possibility of "naked CTCs"
region containing closed timelike curves that are not shielded by a horizon  
(corresponding to the green-shaded area in Fig.~\ref{range_changep}).  
Although the physical realization of such spacetimes is highly speculative and likely disfavored,  
they are permitted within the QIRK parameter space.  
This prompts us to investigate the optical appearance and light propagation characteristics in this region.  
Such an analysis helps us understand how quantum corrections influence light propagation within this region,  
thereby further revealing the impact of quantum effects on causal structure.

In the following discussion,  
we fix the black hole spin parameter to \(a = 1.05\)  
and investigate the effects of quantum corrections by varying the quantum correction parameter \(\widetilde{\omega}\).  
When \(\widetilde{\omega} = 0\),  
this corresponds to the Kerr naked singularity.  
For \(0 < \widetilde{\omega} \le \widetilde{\omega}_m\),  
the solution represents a compact object.

\begin{figure}[!ht]
    \centering
    \includegraphics[width = 0.45 \linewidth]{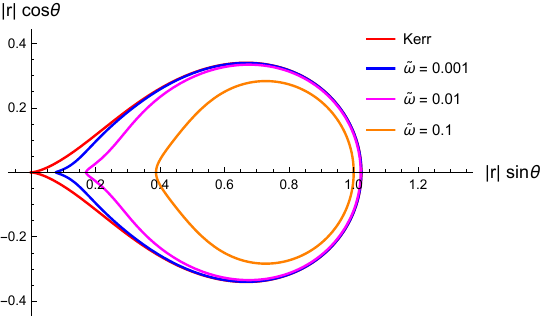}
    \caption{The cross section at \(\phi=0\) of the region where closed timelike curves exist, with the boundary defined by \(g_{\phi\phi}=0\).  
    Here we fix the spin parameter at \(a = 1.05\) and set \(p = 4\).  
    Inside this boundary, \(g_{\phi\phi}<0\), indicating the presence of closed timelike curves along the \(\phi\) coordinate lines.  
    As the quantum correction parameter \(\widetilde{\omega}\) increases, this region steadily shrinks and retreats away from \(r=0\).  
    }
    \label{CTCRegion}
\end{figure}

As shown in Fig.~\ref{CTCRegion},  
the \(\phi=0\) cross section displays the boundary defined by \(g_{\phi\phi}=0\) in both the Kerr spacetime and QIRK spacetimes with varying quantum correction strength \(\widetilde{\omega}\).  
Since we only consider closed timelike curves along the \(\phi\) coordinate lines,  
the region inside this boundary corresponds to where such curves exist.  
In the Kerr case (red curve), the region in which \(g_{\phi\phi}\) becomes negative typically reaches the ring singularity.
However,  
when quantum corrections are introduced, this picture changes significantly.  
Even a small quantum correction (e.g., \(\widetilde{\omega} = 0.001\), blue curve)  
causes the CTC region's boundary to shift slightly outward, away from \(r=0\), compared to Kerr.  
As \(\widetilde{\omega}\) increases,  
the CTC region moves outward and shrinks overall, and this trend becomes more pronounced.  
When \(\widetilde{\omega}\) is large enough (orange curve),  
the CTC region is substantially reduced, and its boundary is clearly distanced from \(r=0\).  
Thus, we see that the inclusion of quantum corrections  
not only regularizes the ring singularity,  
but also significantly modifies the geometry in the immediate neighbourhood of \( r = 0\) (the Kerr ring singularity).

\begin{figure}[!ht] 
    \centering
    \includegraphics[width=0.6 \linewidth]{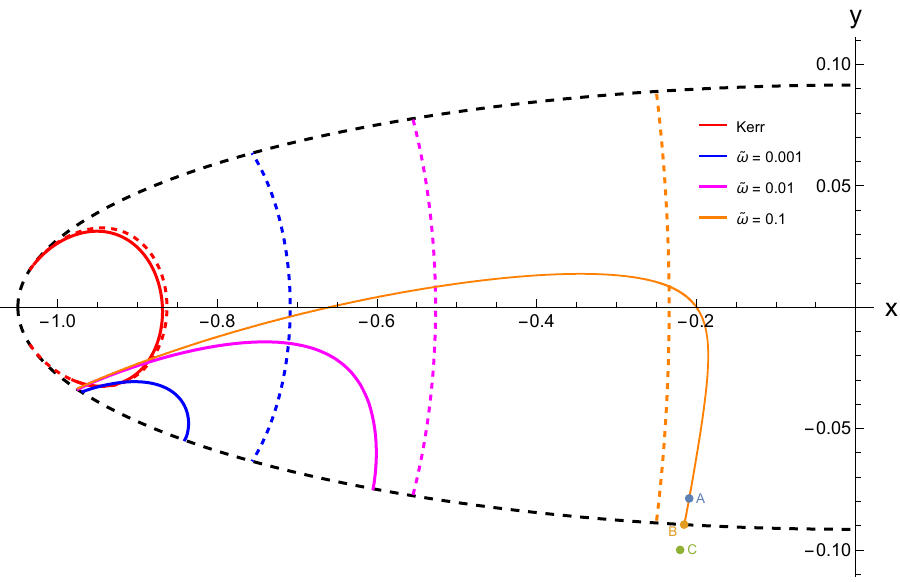}
    \caption{
        Images of CTCs located at \(r = -0.5\) and \(\theta = 80^\circ\) are shown for various spacetime parameters.  
        The black dashed curve marks the turning point at \(r = 0\).  
        Light emitted from \(r < 0\) can only reach the region within this boundary.
        Solid curves represent the direct image and the lensed ring associated with CTCs.  
        Dashed curves correspond to the related photon ring features.  
        Different colors indicate different spacetime models:
        red corresponds to the Kerr spacetime,  
        blue to the QIRK spacetime with \(\widetilde{\omega} = 0.001\),  
        magenta to \(\widetilde{\omega} = 0.01\),  
        and orange to \(\widetilde{\omega} = 0.1\).  
    }
    \label{CTCMainImage}
\end{figure}

\begin{figure}[!ht] 
    \centering
    \includegraphics[width=0.4 \linewidth]{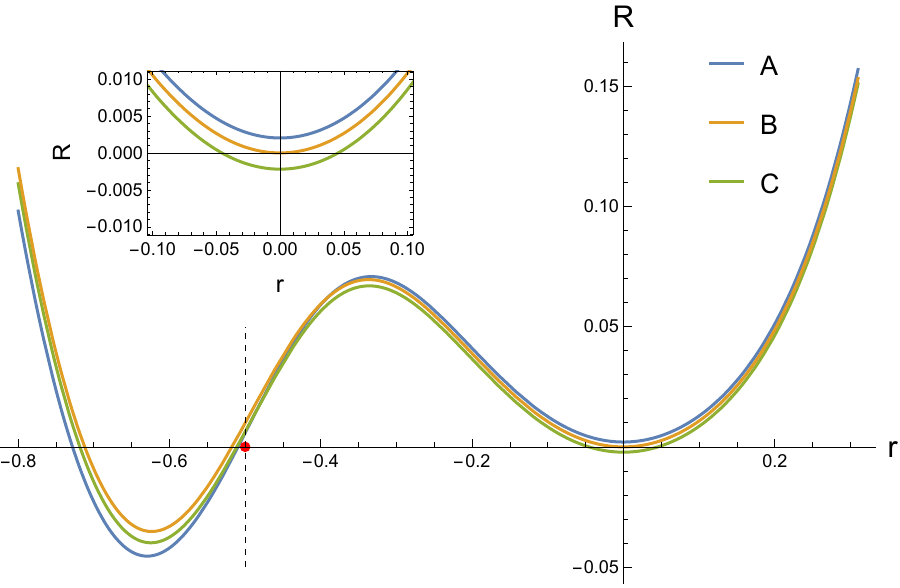}
    \caption{The plot shows the radial function \(R(r)\) as a function of \(r\),  
    with points A, B, and C corresponding to those marked in Fig.~\ref{CTCMainImage}. 
    It can be seen that for light rays reaching points outside the black dashed curve in Fig.~\ref{CTCMainImage},  
    the turning points lie in the region \(r > 0\),  
    which prevents the image from extending beyond the dashed boundary.  
    The black dashed line and red point mark the location of the closed timelike curves at \(r = -0.5\).  
    }
    \label{effBoundaryO}
\end{figure}

Having identified the CTC region under different quantum corrections,  
we now turn to examining how these corrections influence the observed images of light emitted from the region.  
To this end,  
we consider a luminous curve located at \(r = -0.5\), \(\theta = 80^\circ\), and \(t = 0\),  
which lies along a closed timelike curve inside the previously discussed CTC region. 
Fig.~\ref{CTCMainImage} presents the image of this source as seen by a distant observer  
at \(r = 100\), \(\theta_{o} = 85^\circ\), and \(\phi = 0\).  
The black dashed curve consists of points with \(\eta = 0\) for which $R(r) = R'(r) = 0$.
It is not hard to find that $R'(0)$ always  vanishing for any $\tilde{\omega}$, $M$, $a$, $\eta$, and $\xi$. 
Besides, the condition $R(0)=0$ is independent of $\tilde{\omega}$.
Therefore, as shown in the figure, the position of the black dashed curve is independent of $\tilde{\omega}$.
Any light reaching the observer’s screen that was emitted from the CTC region must lie inside this boundary.  
This is confirmed in Fig.~\ref{effBoundaryO}, where a point outside the black dashed curve, such as point C,  
has a turning point at \(r>0\), indicating that such light cannot travel from \(r<0\) to the observer. 
Inside this dashed boundary,
solid curves represent both the direct and lensed images of the source,  
while dashed curves mark the photon ring image that originate from the unstable spherical orbits lying at \( r < 0\) (present in both the Kerr naked singularity and the QIRK compact object).
The corresponding impact parameters for these orbits are those plotted in Fig.\ref{lambdaeta}.

In the Kerr naked singularity,  
the direct image and lensed image (solid line) is relatively compact and confined.  
Even with a small quantum correction in the QIRK model,  
the direct image on the observer's screen shows a modest outward expansion,
and the expansion grows with \(\widetilde{\omega}\).
For \(\widetilde{\omega}=0.1\),  
the direct image extends noticeably further than in the Kerr case,  
indicating that the quantum corrected geometry allows light from the negative-\(r\) region to reach the observer over a wider angle.  
Meanwhile, the photon ring structure (dashed lines) also changes significantly with increasing \(\widetilde{\omega}\).  
In the Kerr case,  
the photon ring associated with the \(r=-0.5\) source forms a relatively tight boundary.  
In QIRK spacetimes,  
this curve shifts outward and becomes flatter as \(\widetilde{\omega}\) increases.  
This behavior suggests that quantum corrections not only affect the propagation of directly emitted light,  
but also fundamentally modify the unstable photon orbits near \(r=0\).  
Taken together,  
the changes in both direct and higher-order images show how the QIRK model's regularization of the \(r=0\) region alters the observable features of nearby sources.

\begin{figure}[htbp]
    \begin{minipage}{0.45 \textwidth}
        \centering
        \includegraphics[width=\linewidth]{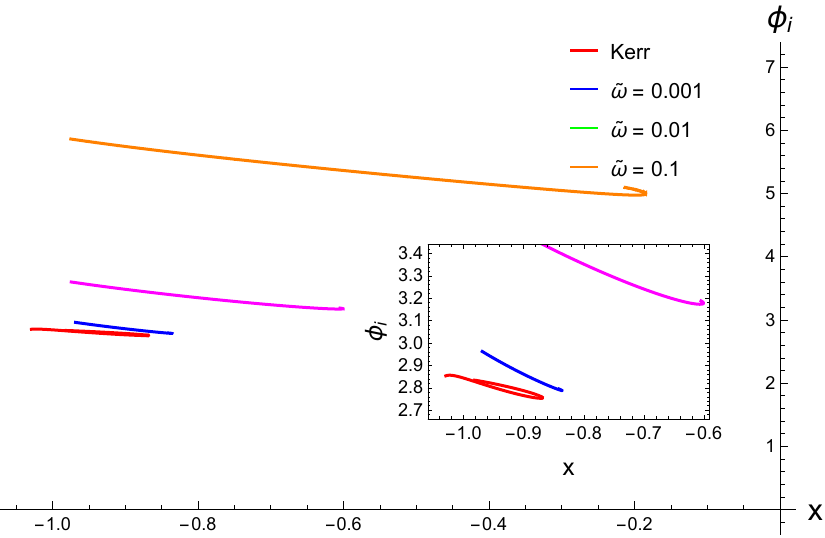}
        \end{minipage}
        \hfill
        \begin{minipage}{0.45 \textwidth}
        \centering
        \includegraphics[width=\linewidth]{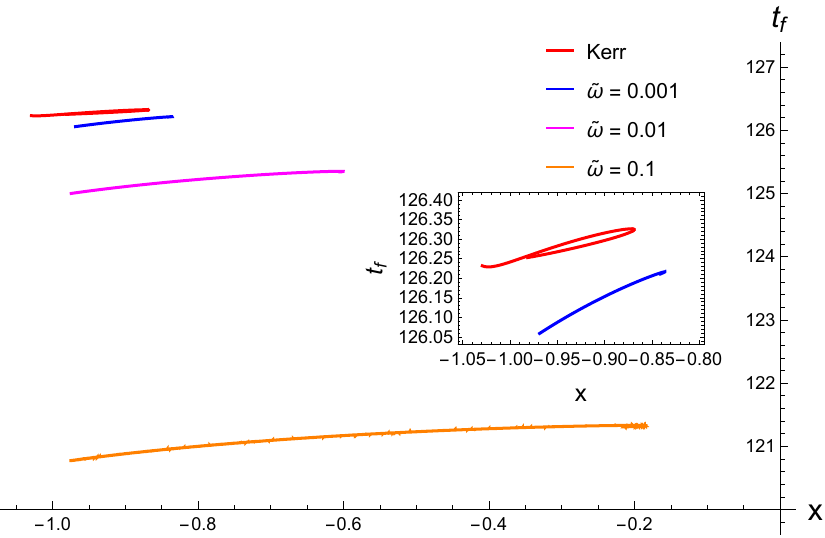}
        \end{minipage}
        \caption{
            Left panel: Initial azimuthal angle \(\phi_i\) of photons emitted from the CTCs as a function of the observed horizontal coordinate \(x\). 
            Right panel: Arrival time \(t_f\) of these photons at the observer as a function of \(x\). 
            Different colors correspond to different spacetime parameters as shown in Fig.~\ref{CTCMainImage}.}
\label{CTCPhiT}
\end{figure}

\begin{figure}[htbp]
    \begin{minipage}{0.45 \textwidth}
        \centering
        \includegraphics[width=\linewidth]{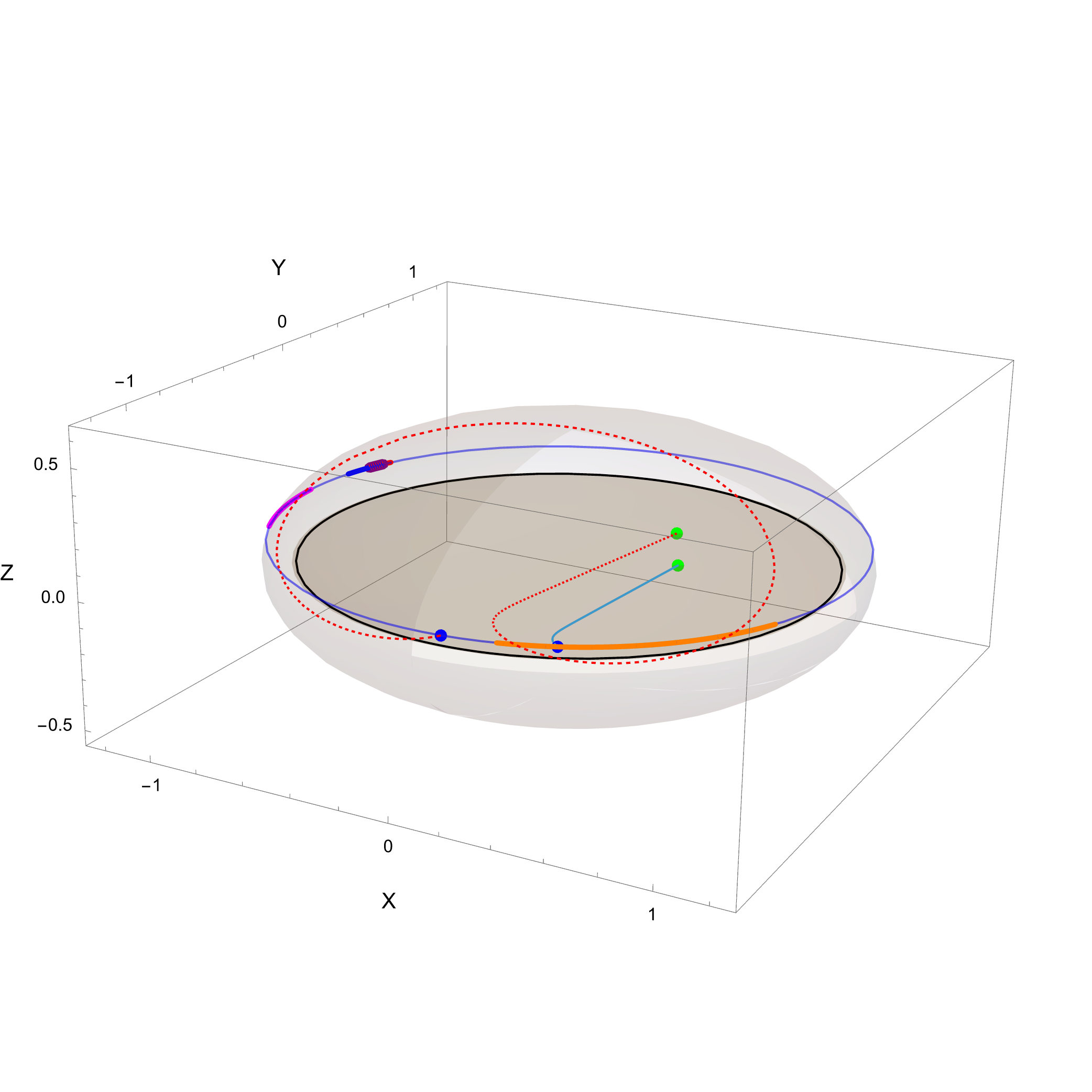}
        \end{minipage}
        \begin{minipage}{0.45 \textwidth}
        \centering
        \includegraphics[width=\linewidth]{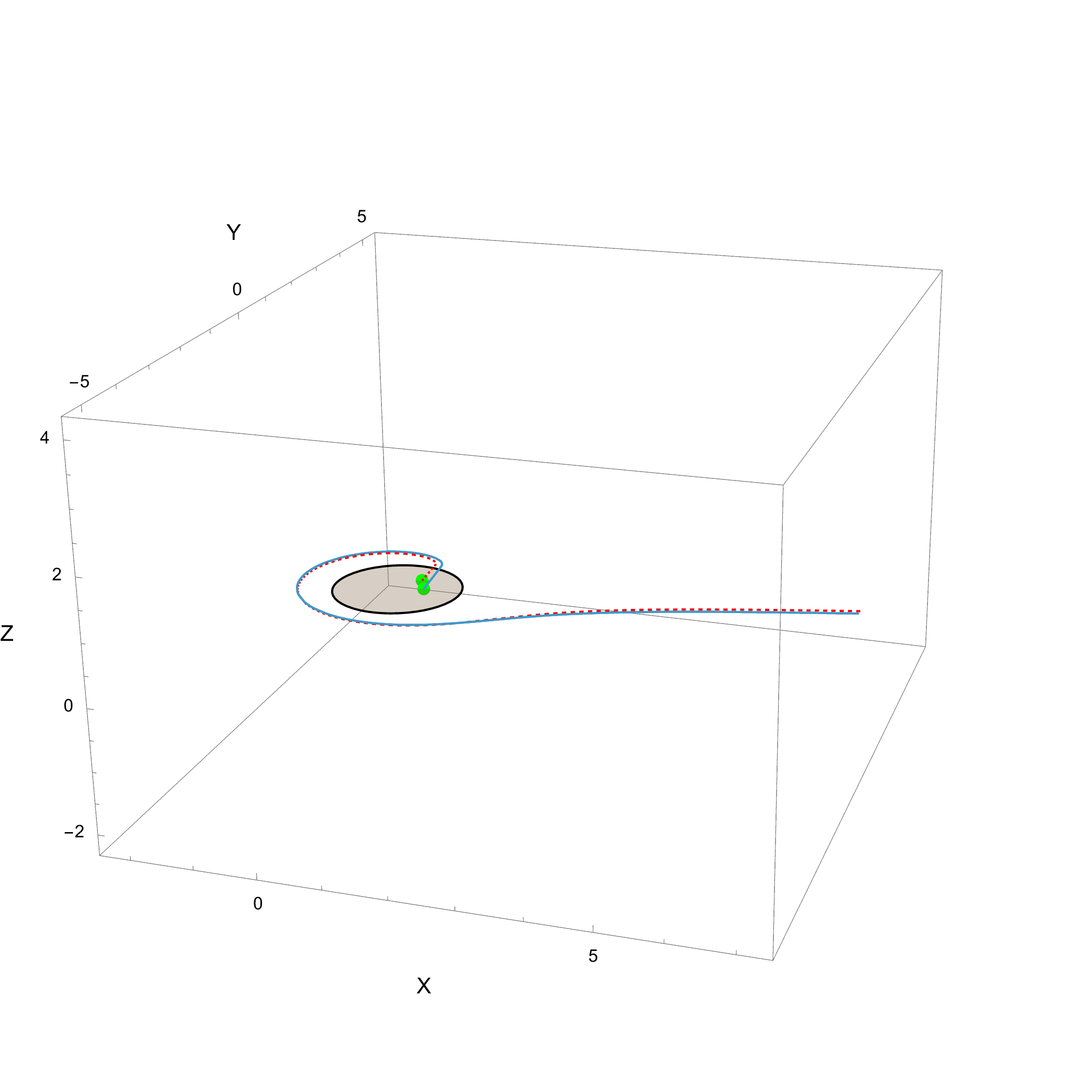}
        \end{minipage}
        \caption{
        Left panel: Two bundles of rays originate from a closed timelike curve at \(r = -0.5\), \(\theta_s = 80^\circ\).
        The solid curves show directly emitted photons, while the red dashed curve traces the rays that form the photon ring image.
        Along the CTC (light-blue), the red, blue, magenta, and orange segments indicate the region visible to the observer for the Kerr case and for \(\widetilde{\omega} = 0.001\), \(0.01\), and \(0.1\), respectively--visibility here refers only to the direct and lensed images.
        Right panel: These rays cross the equatorial disk at \(r = 0\) (dark gray) and propagate outward to the observer; the black circle indicates the disk's outer boundary.
        }
\label{trajectoryPlot}
\end{figure}

The left panel of Fig.~\ref{CTCPhiT} shows the relation between the initial emission angle \(\phi_i\) at the source and the horizontal screen coordinate \(x\) for the direct and lensed images (solid curves in Fig.\ref{CTCMainImage}).
As \(\widetilde{\omega}\) increases, the range of \(x\) expands, consistent with the changes in Fig.~\ref{CTCMainImage}.  
For any fixed \(x\), the required emission angle \(\phi_i\) shifts significantly with \(\widetilde{\omega}\), indicating that quantum corrections alter the mapping from source angle to image position.  
This shift arises from modified null geodesics in the QIRK spacetime and reflects changes in gravitational lensing for \(r<0\).  
Each curve also develops a small "hook" at its right end, indicating that one emission angle \(\phi\) can produce two observed \(x\) values—this corresponds to the lensed image. 
Moreover, the observer cannot receive photons emitted at all \(\phi\) angles.  
The range and location of observable \(\phi\) angles grow with \(\widetilde{\omega}\), moving toward \(2\pi\).

Figure~\ref{trajectoryPlot} illustrates two families of null rays emitted from the CTC at \(r=-0.5\) and traversing the \(r=0\) surface before reaching the observer.
The solid curves trace the directly emitted photons, 
while the red dashed curves correspond to rays that form the photon ring image.
In the left panel of Fig.~\ref{trajectoryPlot}, 
the red, blue, magenta, and orange segments mark the emission angles visible to the observer for \(\widetilde{\omega}=0\), \(0.001\), \(0.01\), and \(0.1\), respectively, matching the behavior seen in Fig.~\ref{CTCPhiT}.  
The region of observable emission angles also depend on the source radius \(r_{s}\),  
the source polar angle \(\theta_{s}\),  
and the observer's coordinates \(r_{o}\) and \(\theta_{o}\).  
Here we hold both the source and the observer at fixed coordinates \((r, \theta)\) while examining the effects of quantum corrections.

It is worth noting that we have omitted the photon ring image here: the photon ring probes only the underlying \( r< 0 \) geometry and arises from unstable spherical orbits that can be launched from any azimuthal angle \( \phi_i\),
so it does not carry additional information about the closed timelike curve itself. 
For the same reason, our subsequent analysis of photon arrival times \( t_f \) focuses exclusively on the direct and lensed images (the solid curves in Fig.~\ref{CTCMainImage}).

The right panel of Fig.~\ref{CTCPhiT} shows the arrival time \(t_f\) of these photons at the distant observer as a function of their observed horizontal coordinate \(x\).  
For QIRK spacetimes with small quantum corrections, 
the arrival times shift only slightly compared to the Kerr case, and the shape of the curve remains largely unchanged.  
For larger quantum corrections, the arrival time curve changes markedly: 
photons arrive earlier than in the other cases.  
At the right end of each curve a hook-like feature appears, 
corresponding to the lensed image arrival times shown in the left of Fig.~\ref{CTCPhiT}.  
For example, for $\widetilde{\omega}=0.1$, all points on the CTC emit rays simultaneously at $t=0$, and these rays are observed at different time by the observer.
A bright spot firstly appears at the left endpoint of the orange curve.
Then it moves to the right in the image. 
At the time corresponding to the other end of the orange curve, another bright spot appears and moves to left. 
Eventually, the two bright spots converge and vanish.
The clear shift in arrival times with increasing \(\widetilde{\omega}\) highlights the strong influence of quantum corrections on the underlying spacetime geometry.

\section{Conclusion}\label{Conclusion}

In this work, we focus on the basic properties of the QIRK black hole,  
including its parameter space,
its observational signatures in thin disk imaging,  
parameter constraints from EHT data,  
and a preliminary exploration of the region containing closed timelike curves.

First, by analyzing the conditions for spacetime regularity,  
horizon existence, and the absence of CTCs,  
we identified the allowed parameter range of \((a, \widetilde{\omega})\).  
To ensure regularity, the exponent \(p\) in the running Newton constant \(G(r)\) must be an even integer greater than three.  
Horizon existence requires \(\widetilde{\omega} \le \widetilde{\omega}_{c}(a)\),  
while avoiding CTCs demands \(\widetilde{\omega} > \widetilde{\omega}_{m}(a)\).  
Together these conditions define the physically viable region in Fig.~\ref{range_changep},
with the maximum spin satisfying \(a_{\text{max}} < 1\).

Next, we examine the optical appearance of a QIRK black hole illuminated by a geometrically thin, optically thick accretion disk.
We find that the quantum correction \(\widetilde{\omega}\) has only a small effect on the intensity and redshift distributions;
its main impact is a reduction of the overall image brightness.  
Thus, separating QIRK from Kerr based solely on image morphology or brightness will be challenging and likely requires very sensitive measurements.

We then use shadow observables—the angular diameter \(\theta_{d}\) and circularity deviation \(\Delta C\)—  
together with EHT observations of M87* and Sgr~A*,  
to constrain \((a, \widetilde{\omega})\).  
These constraints form a narrow band in parameter space.  
For \(\widetilde{\omega}=0\), they recover known limits on the spin \(a\) for Kerr.  
As \(\widetilde{\omega}\) grows, the allowed range of \(a\) shrinks.  
Interestingly, even nearly nonrotating QIRK black holes (\(a\approx0\)) can match Sgr~A* data  
if \(\widetilde{\omega}\) lies close to its critical value \(\widetilde{\omega}_{c}(0)\).

Finally, we conduct a preliminary study of the optical characteristics of region associated with closed timelike curves in QIRK spacetime.   
Quantum corrections alter light paths in this region:  
the direct image becomes asymmetric,  
and both direct and secondary images expand outward as \(\widetilde{\omega}\) increases.  
Photon arrival times also shift, highlighting the effect of quantum corrections on deep interior geometry.

In summary, the QIRK model offers a regular alternative to Kerr with distinctive observational features.  
Although some effects are subtle, variations in image brightness and EHT constraints provide key tests of this quantum gravity–inspired solution.  
Future high resolution observations and multiwavelength data should tighten limits on \(\widetilde{\omega}\) and \(a\).  
On the theoretical side,  
complex astrophysical environments (e.g., thick accretion disks and jets)  
and detailed analyses of thermodynamic properties, stability, and quasinormal modes \cite{Stashko:2024wuq}  
are important aspects for future studies.

\section*{Acknowledgement}
We would like to thank Nobuyoshi Ohta and Chiang-Mei Chen for their useful discussions and communications. 
This work was supported in part by the National Natural Science Foundation of China with grants No.12075232 and
No.12247103.

\bibliographystyle{unsrt}
\bibliography{main}

\end{document}